\documentclass[lettersize,journal]{IEEEtran}
\usepackage{amsmath,amsfonts}
\usepackage{array}
\usepackage{textcomp}
\usepackage{stfloats}
\usepackage{url}
\usepackage{verbatim}
\usepackage{graphicx}
\usepackage{cite}

\usepackage{tabularx}
\usepackage{booktabs}
\usepackage{xcolor}

\usepackage{multirow}
\usepackage{diagbox}
\usepackage{makecell}

\usepackage{algpseudocode}
\usepackage[ruled,vlined]{algorithm2e}
\let\oldnl\nl
\newcommand{\nonl}{\renewcommand{\nl}{\let\nl\oldnl}}

\usepackage{tablefootnote}
\usepackage{threeparttable}

\usepackage{caption}
\usepackage[labelformat=simple]{subcaption}

\renewcommand{\dot}[1] {\overset{\,_{\mbox{\Large .}}}{#1}} 

\usepackage{enumitem}
\setlist[itemize]{leftmargin=2ex}

\usepackage{pifont} 
\usepackage{xcolor}
\definecolor{darkgreen}{rgb}{0,0.8,0}
\definecolor{darkred}{rgb}{0.8,0,0}
\newcommand{\greencheck}{{\color{darkgreen}\ding{51}}}
\newcommand{\redcross}{{\color{darkred}\ding{55}}}

\begin{document}

\title{CHIP: Chameleon Hash-based Irreversible Passport for Robust Deep Model Ownership Verification and Active Usage Control}

\author{
    {
        Chaohui~Xu,
        Qi~Cui,~\IEEEmembership{Member,~IEEE,}
        and~Chip-Hong~Chang,~\IEEEmembership{Fellow,~IEEE}
    }
    % \thanks{This research is supported by the National Research Foundation, Singapore, and Cyber Security Agency of Singapore under its National Cybersecurity Research \& Development Programme (Development of Secured Components \& Systems in Emerging Technologies through Hardware \& Software Evaluation \texttt{<} NRF-NCR25-DeSNTU-0001 \texttt{>}). Any opinions, findings and conclusions or recommendations expressed in this material are those of the author(s) and do not reflect the view of National Research Foundation, Singapore and Cyber Security Agency of Singapore.}
    \thanks{C. Xu is with the School of Electrical and Electronic Engineering, Nanyang Technological University, Singapore 639798. Q. Cui is with the Engineering Research Center of Digital Forensics, School of Computer Science, Nanjing University of Information Science and Technology, Nanjing, China 210044. C. H. Chang is with the School of Electrical and Electronic Engineering and National Integrated Centre for Evaluation (NiCE), Nanyang Technological University, Singapore 639798. (Email: \{chaohui001@e.ntu.edu.sg, echchang@ntu.edu.sg\}). Corresponding author: C. H. Chang.}
}

\markboth{Journal of \LaTeX\ Class Files,~Vol.~14, No.~8, August~2021}%
{Shell \MakeLowercase{\textit{et al.}}: A Sample Article Using IEEEtran.cls for IEEE Journals}


\maketitle

\begin{abstract}

The pervasion of large-scale Deep Neural Networks (DNNs) and their enormous training costs make their intellectual property (IP) protection of paramount importance. Recently introduced passport-based methods attempt to steer DNN watermarking towards strengthening ownership verification against ambiguity attacks by modulating the affine parameters of normalization layers. Unfortunately, neither watermarking nor passport-based methods provide a holistic protection with robust ownership proof, high fidelity, active usage authorization and user traceability for offline access distributed models and multi-user Machine-Learning as a Service (MLaaS) cloud model. In this paper, we propose a Chameleon Hash-based Irreversible Passport (CHIP) protection framework that utilizes the cryptographic chameleon hash function to achieve all these goals. The collision-resistant property of chameleon hash allows for strong model ownership claim upon IP infringement and liable user traceability, while the trapdoor-collision property enables hashing of multiple user passports and licensee certificates to the same immutable signature to realize active usage control. Using the owner passport as an oracle, multiple user-specific triplets, each contains a passport-aware user model, a user passport, and a licensee certificate can be created for secure offline distribution. The watermarked master model can also be deployed for MLaaS with usage permission verifiable by the provision of any trapdoor-colliding user passports. CHIP is extensively evaluated on four datasets and two architectures to demonstrate its protection versatility and robustness. Our code is released at https://github.com/Dshm212/CHIP.

\end{abstract}


\section{Introduction} \label{sec}

Over the past decade, deep neural network (DNN) parameters have increased exponentially by five orders of magnitude, which make training a model from scratch extremely time consuming and costly~\cite{achiam2023gpt,bai2023qwen,touvron2023llama}. The process of developing a new DNN model involves extensive data collection, precise labeling, substantial computational resources, and expert knowledge. It is no surprise that rigorously trained models have become prime targets for piracy, unauthorized redistribution, and illicit use. Recent studies~\cite{tramer2016stealing,yu2020deepem,shi2018generative,orekondy2019knockoff, kariyappa2021maze,mishra2024too,zheng2024overview} have underscored the severity of DNN model intellectual property (IP) infringement, which call for more versatile, robust and holistic protection.

Various DNN watermarking methods~\cite{uchida2017embedding,chen2019deepmarks,nie2024deep,lv2023robustness,wang2021riga,wang2023free,adi2018turning,zhang2018protecting} have emerged to embed ownership marks into the model by modifying network weights or adjusting decision boundaries to specific inputs (triggers) with minimal or no degradation on the primary task performance. Though many of these approaches can achieve black-box ownership verification with robust watermark against removal and modification, they are susceptible to ambiguity attacks, wherein attackers embed an additional watermark to claim ownership.

\begin{table}[t]
    \caption{Qualitative comparison of passport-based IP protection methods. \greencheck indicates presence, and \redcross indicates absence.}
    \centering
    \renewcommand\arraystretch{1.1}
    \resizebox{0.95\linewidth}{!}{
      \begin{tabular}{l|ccccc}
        \Xhline{2\arrayrulewidth}
        \multirow{2}{*}{Method}                                            & \multicolumn{1}{c|}{\multirow{2}{*}{Watermark}} & \multicolumn{1}{c|}{\multirow{2}{*}{Fidelity}} & \multicolumn{1}{c|}{\multirow{2}{*}{\begin{tabular}[c]{@{}c@{}}Enhanced\\ Robustness\end{tabular}}} & \multicolumn{2}{c}{Multi-user Control}            \\ \cline{5-6} 
                                                                           & \multicolumn{1}{c|}{}                           & \multicolumn{1}{c|}{}                             & \multicolumn{1}{c|}{}                                                                            & \multicolumn{1}{c|}{Online} & Offline                    \\ \Xhline{2\arrayrulewidth}
        DeepIPR~\cite{fan2019rethinking,fan2021deepipr}                    & \greencheck                      & \redcross                        & \redcross                                                                       & \redcross  & \redcross \\
        PAN~\cite{zhang2020passport}                                       & \greencheck                      & \greencheck                        & \redcross                                                                       & \redcross  & \redcross \\
        TdN~\cite{liu2023trapdoor}                                         & \greencheck                      & \greencheck                        & \greencheck                                                                       & \redcross  & \redcross \\
        SteP~\cite{cui2024steganographic}                                  & \greencheck                      & \greencheck                        & \greencheck                                                                       & \greencheck  & \redcross \\
        CHIP (Ours)                                                        & \greencheck                      & \greencheck                        & \greencheck                                                                       & \greencheck  & \greencheck \\ \Xhline{2\arrayrulewidth}
      \end{tabular}
    }
    \label{tab:feature_comparison}
    \vspace{-5mm}
\end{table}

To resolve this copyright conflict, Fan et al.~\cite{fan2019rethinking} proposed the first passport-based watermarking method, which replaces selected normalization (henceforth abbreviated as norm) layers in the target model with specially designed passport layers. High inference performance, similar to that of an unprotected model, can be achieved only when the correct passport features is present in these layers. Thereafter, several advanced passport-based methods~\cite{fan2021deepipr,zhang2020passport,liu2023trapdoor,cui2024steganographic} with enhanced robustness and flexibility have been introduced. However, existing passport-based methods still have limitations, including reduced performance fidelity~\cite{fan2019rethinking,fan2021deepipr}, poor robustness against stronger ambiguity attacks with oracle passports~\cite{fan2019rethinking,fan2021deepipr,zhang2020passport}, trading signature privacy for enhanced robustness~\cite{liu2023trapdoor}. More importantly, all these methods do not support single model deployment for Machine Learning as a Service (MLaaS) with authorized access control of multiple users and ad hoc user subscription and withdrawal. This problem was solved in~\cite{cui2024steganographic} at the expense of limiting active usage control and traceability on offline distributed instances.

This paper introduces Chameleon Hash-based Irreversible Passport (CHIP), a new versatile IP protection framework that overcomes the limitations of existing passport-based methods. The provably secure cryptographic chameleon hash function is utilized to create an \textbf{immutable} signature from the owner passport and licensor certificate to watermark the master model. The trapdoor-collision property of chameleon hash allows the model owner to generate multiple user models based on the master model for offline distribution, without compromising robustness against ambiguity attacks or requiring extensive model retraining. Each user model is bound to a distinct user passport and a licensee certificate. In addition, a skip connection is introduced to the passport layer to create strong dependence between critical affine factors and the passport. This architectural enhancement guarantees that each user model remains operational only with its designated paired user passport. Consequently, the model owner can actively restrict the usage of user models exclusively to authorized users possessing the valid passports. Furthermore, CHIP also allows the model owner to establish ownership proof and actively trace registered users for unauthorized use or resale of distributed models. The intended chameleon signature collision to the immutable signature can only be produced by its registered user with the designated paired passport and licensee certificate issued by the model owner. CHIP can also be applied to the online MLaaS mode with access control and traceability of a large number of registered users by the design of collision-resistant chameleon hash. Table~\ref{tab:feature_comparison} provides a qualitative comparison of attributes across different passport-based IP protection methods. Our contributions are as follows:
\begin{itemize}
    \item We propose CHIP, a chameleon hash-based DNN IP protection method which effectively and efficiently achieves not only model watermarking but also multi-user active control in both online and offline scenarios.
    \item Through extensive evaluations on four datasets and two model architectures, we demonstrate the superior performance of CHIP over existing passport-based methods in terms of effectiveness, fidelity, and robustness. The watermark can be successfully embedded into the target model with no or negligible accuracy degradation. CHIP is also resistant to various ambiguity attacks and removal attacks.
    \item We verify CHIP's capability for active control in both online and offline deployment modes.
    \item Beyond image classification, we validate the effectiveness of CHIP on graph classification to showcase its versatility on diverse ML tasks.
\end{itemize}

The rest of this paper is structured as follows. Section~\uppercase\expandafter{\romannumeral2} reviews related works. Section~\uppercase\expandafter{\romannumeral3} introduces our threat model, provides background knowledge on chameleon hash, and discusses technical details of existing passport-based methods. The proposed CHIP is elaborated in Section~\uppercase\expandafter{\romannumeral4}, followed by experimental results and analysis in Section~\uppercase\expandafter{\romannumeral5}. The paper is concluded in Section~\uppercase\expandafter{\romannumeral6}.

\section{Related Works}

DNN models are facing security and privacy threat to model stealing attacks~\cite{zheng2024overview} that aim to either precisely stealing crucial components of the target model~\cite{tramer2016stealing,yu2020deepem,mishra2024too} or creating a substitute model that has the same or approximate functionality as the target model~\cite{shi2018generative,orekondy2019knockoff, kariyappa2021maze}. Protection of DNN against IP theft and related security threats can be broadly categorized into passive and active protection methods.

DNN watermarking achieves passive IP rights protection by concealing the copyright information into the target model for verification. The first DNN watermarking method~\cite{uchida2017embedding} embeds secret information into the model's weights by including an additional regularization loss during training to constrain the biases of the embedded hidden layers to follow a particular distribution. Following this line of thought, advanced DNN watermarking techniques further enhance the robustness~\cite{chen2019deepmarks,nie2024deep,lv2023robustness}, performance fidelity~\cite{wang2021riga,wang2023free}, transferability~\cite{adi2018turning}, and generalizability~\cite{zhang2018protecting}. Instead of hiding extraneous information into the network, DNN fingerprinting methods extract unique intrinsic characteristics from pretrained model for ownership proof without modifying the model parameters. Specifically, proprietary training details~\cite{jia2021proof}, decision boundaries~\cite{cao2021ipguard,pan2022metav,ren2023ganfinger,zhuang2024deepreg}, and carefully selected distinctive weights~\cite{zheng2022dnn} have been explored to create unique fingerprints to identify originally designed and trained DNN models.

On the other hand, active protection methods aim to proactively prevent unauthorized access by restricting the model's utility without a valid key. Preemptive control can be achieved through embedding the target model into a trustworthy hardware~\cite{chakraborty2020hardware, lin2024older}, encrypting the target model~\cite{lin2020chaotic,zhou2023nnsplitter,wong2024snngx}, and training the target model on processed data and/or with carefully-designed algorithms~\cite{wang2021non,ren2022protecting, tang2023deep,li2024securenet}.

Originated with the aim against ambiguity attacks, a distinct class of protection methods~\cite{fan2019rethinking,fan2021deepipr,zhang2020passport,liu2023trapdoor,cui2024steganographic} replaces selected normalization layers in the target model with purpose-designed passport layers to embed the watermark for anti-forgery ownership identification. Depending on the provision, these methods can be passive or active. The proposed CHIP belongs to the active category of passport-based methods. Section~\ref{sec:existing_passport} provides a comprehensive review and analysis of existing passport-based approaches.

\section{Preliminaries}

\subsection{Threat Model}

The owner is assumed to have complete knowledge and full control of the training pipeline to embed watermark (aka signature in passport-based methods) for attesting the ownership of an infringed model and identifying its registered users. The watermark should be robust against potential attacks, such as removal attacks and ambiguity attacks that forge watermarks for false ownership claims.

A malicious registered user who has access to the protected model and genuine user passport may also alter the model to remove the watermark or counterfeit the passport to create a copyright conflict. The attacker has limited training data available, and this deed must not unduly degrade the model's performance.

\subsection{Chameleon Hash}

Hash functions (MD5~\cite{rivest1992md5}, SHA-1~\cite{eastlake2001us}, etc.) are commonly used in digital signature schemes due to their one-wayness and collision resistance. Given a message $m$, it is computationally infeasible to find another collision message $m^\prime \neq m$ such that $\text{Hash}(m) = \text{Hash}(m^\prime)$.

Chameleon hash~\cite{krawczyk1998chameleon} is a special trapdoor hash function that provides controlled flexibility in generating collisions. Let $\mathcal{PK}$ and $\mathcal{SK}$ be the paired public key and secret key, respectively. The chameleon hash value is computed as $h = \operatorname{CH}(\mathcal{PK}, m, r)$, where $m$ and $r$ are a message and a random string, respectively. The chameleon hash function has the same collision-resistant property as traditional hash functions if only the public key $\mathcal{PK}$ is available. Generating collisions with only $\mathcal{PK}$ is computationally intractable as the chameleon hash is designed based on the hard discrete mathematical problem~\cite{mccurley1990discrete}. However, when the secret key $\mathcal{SK}$ is known, collisions of an arbitrary distinct message $m^\prime \neq m$ can be easily achieved. A corresponding random string $r^\prime$ can be efficiently found by trapdoor collision $r^\prime=\operatorname{Col}(\mathcal{SK}, m, r, m^\prime)$ that satisfies $\operatorname{CH}(\mathcal{PK}, m, r) = \operatorname{CH}(\mathcal{PK}, m^\prime, r^\prime)$. More details about chameleon hash are given in the Supplementary Material, Sec. S1.

We use chameleon hash to create \textbf{mutable} user passports by trapdoor collisions to the owner passport (message) while maintaining the \textbf{immutable} signature (hash value).

\begin{figure}[tbh]
    \centering
    \begin{subfigure}{0.85\linewidth}
        \centering
        \includegraphics[width=\textwidth]{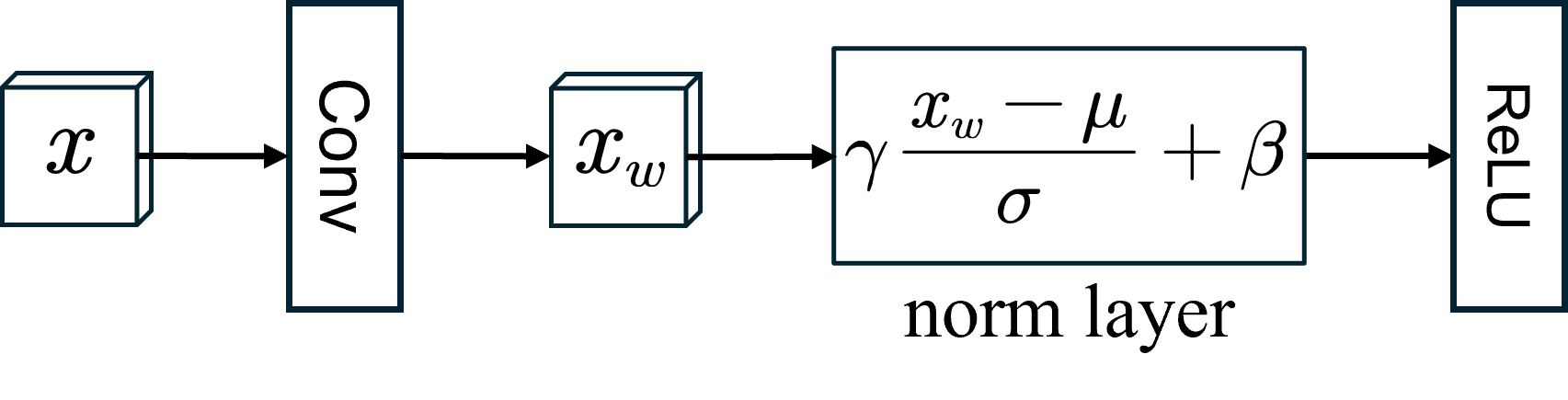}
        % \vspace{-3mm}
        \caption{A typical convolutional block with a norm layer.}
        \label{fig:conv_norm}
    \end{subfigure}
    \\
    \vspace{2mm}
    \begin{subfigure}{0.90\linewidth}
        \centering
        \includegraphics[width=\textwidth]{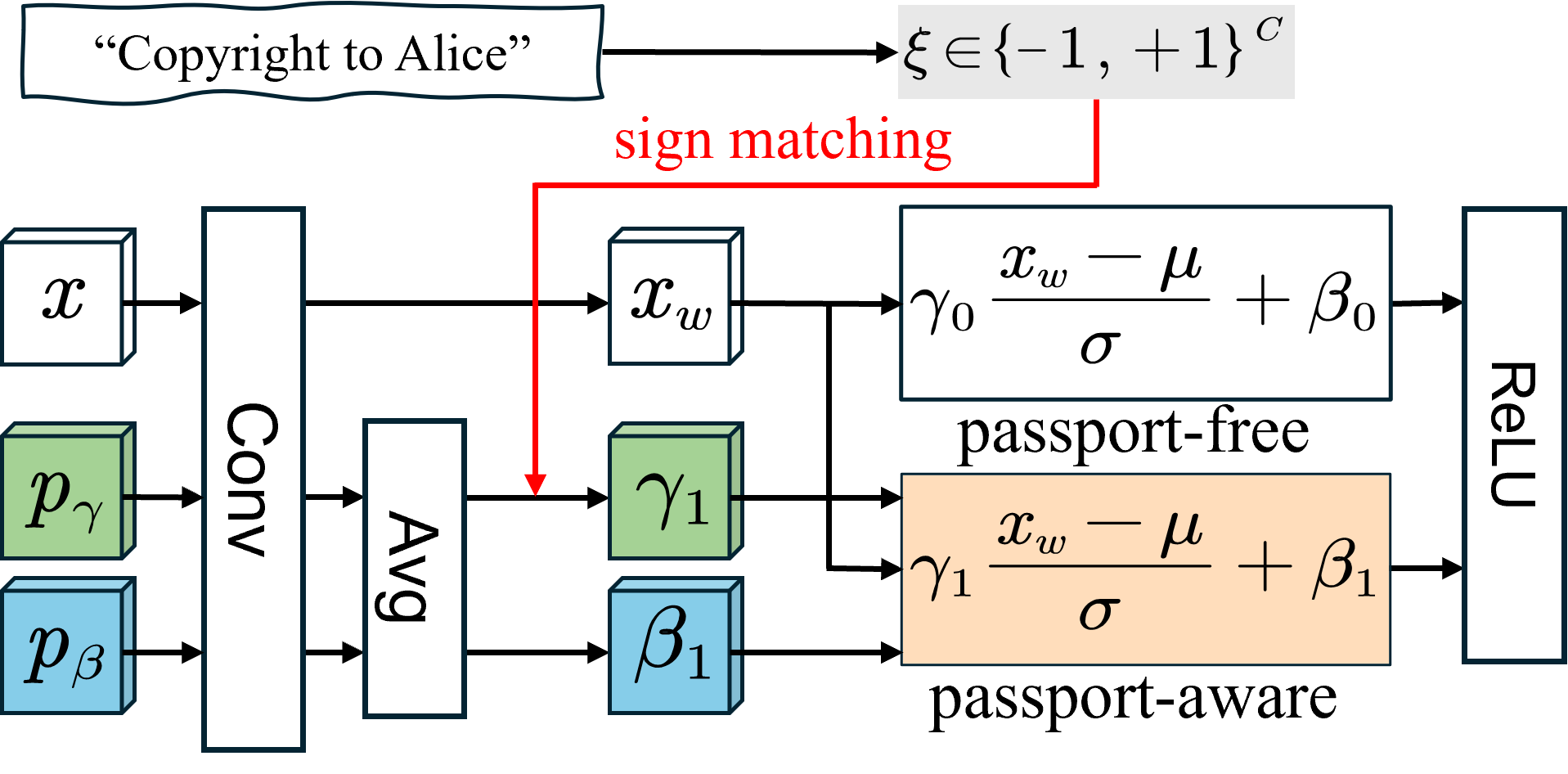}
        % \vspace{-3mm}
        \caption{Passport layer of DeepIPR.}
        \label{fig:passport_layer_DeepIPR}
    \end{subfigure}
    
    \caption{Structures of (a) a typical convolutional block, and (b) the passport layer of DeepIPR.}
    \vspace{-5mm}

\end{figure}

\subsection{Existing Passport-based Protections Methods} \label{sec:existing_passport}

Normalization (norm) layers are extensively used in deep models to improve training efficiency and enhance performance. Let $x$ denotes the input feature map and $\otimes$ be the convolution operator. As shown in Fig.~\ref{fig:conv_norm}, $x$ is first convoluted by the trainable convolutional kernel $\mathbf{W}$ to $x_w = \mathbf{W} \otimes x$, and then further normalized to:
\begin{equation}
  \hat{x} = \gamma \frac{x_w - \mu}{\sigma} + \beta,
  \label{eq:norm_operation}
\end{equation}
where $\mu$ and $\sigma$ are the mean and standard deviation (std) of $x_w$, respectively. Convoluted feature maps are calculated differently for different normalization methods. For example, Batch Normalization (BN)~\cite{ioffe2015batch} computes $\mu$ and $\sigma$ over mini-batches during training, while Group Normalization (GN)\cite{wu2018group} divides features into groups and calculates $\mu$ and $\sigma$ on-the-fly during inference. $\gamma$ and $\beta$ are the affine scale and bias factors. They play a crucial role in projecting normalized features to appropriate scales.

\subsubsection{DeepIPR~\cite{fan2019rethinking,fan2021deepipr}}

This is the first passport-based DNN IP protection scheme. It replaces selected convolutional blocks in the target model with passport layers to embed a robust watermark as shown in Fig~\ref{fig:passport_layer_DeepIPR}. Given a passport $p = \{p_\gamma, p_\beta\}$ consisting of pre-defined feature maps $p_\gamma$ and $p_\beta$, that shares the same spatial dimensions as the input $x$, the passport layer operates through two norm branches:
\begin{equation}
  \hat{x}=
  \begin{cases}
        \gamma_0 \frac{x_w - \mu}{\sigma} + \beta_0,     \qquad & \text{passport-free},\\
        \gamma_1 \frac{x_w - \mu}{\sigma} + \beta_1,            & \text{passport-aware},
  \end{cases}
\end{equation}
where the upper \textbf{passport-free} branch contains two learnable affined factors $\gamma_0$ and $\beta_0$ that are trained originally without the passport, while the lower \textbf{passport-aware} branch utilizes feature maps convoluted from the passport as the affine factors:
\begin{equation}
    \gamma_1 = wp_\gamma, \quad \beta_1 = wp_\beta,
\end{equation}
where $wp_\gamma = \operatorname{Avg}(\mathbf{W} \otimes p_\gamma)$ and $wp_\beta = \operatorname{Avg}(\mathbf{W} \otimes p_\beta)$, with $\operatorname{Avg}(\cdot)$ being the average pooling function. The statistics mean and std are shared by the two branches.

The target model $\mathcal{M}$ is trained on the training dataset $D_\text{tr}$ to achieve high performance with both passport-free and passport-aware branches with the following losses:
\begin{equation}
    \begin{aligned}
        & \mathcal{L}_f = \mathbb{E}_{(\mathbf{x}_i, \mathbf{y}_i) \sim D_\text{tr}} \big[\mathcal{L}_\text{CE}(\mathcal{M}^\text{f}(\mathbf{x}_i), \mathbf{y}_i)\big], \\
        & \mathcal{L}_a = \mathbb{E}_{(\mathbf{x}_i, \mathbf{y}_i) \sim D_\text{tr}} \big[\mathcal{L}_\text{CE}(\mathcal{M}^\text{a}(\mathbf{x}_i), \mathbf{y}_i)\big],
    \end{aligned}
    \label{eq:cross_entropy}
\end{equation}
where $\mathcal{L}_\text{CE}$ denotes the cross-entropy loss. $\mathcal{M}^f$ and $\mathcal{M}^a$ denote the model with only the passport-free or passport-aware branch, respectively.

Moreover, the model owner arbitrarily creates a copyright text $\mathcal{T}$ (e.g. ``Copyright to Alice'') and converts it to a $C$-bit $\pm 1$ signature sequence $\xi=\{\xi_1, \xi_2, \cdots, \xi_C\} \in \{-1, +1\}^C$. The signs of $wp_\gamma$ are enforced to match $\xi$ as follows:
\begin{equation}
    \vspace{-2mm}
    \mathcal{L}_\text{s} = \sum^C_{i=1}\text{Max}\big[(\tau - \xi_i \cdot (wp_\gamma)_i), 0\big],
    \label{eq:sign_matching}
\end{equation}
where $\tau$ is a small positive threshold (0.1 in previous works) to keep the magnitudes of $wp_\gamma$ low and thereby its signs are lazy-to-flip during fine-tuning due to small gradient.

By jointly optimized with the three losses ($\mathcal{L}_f$, $\mathcal{L}_a$, and $\mathcal{L}_s$), the convolutional layer of a passport layer is able to: (1) properly extract features from the input $x$; (2) project $p_\gamma$ and $p_\beta$ to the correct affine factors $\gamma_1$ and $\beta_1$; and (3) ensure the signs of $wp_\gamma$ matches the signature string $\xi$.

Upon training, the passport-free model $\mathcal{M}^\text{f}$ is distributed to users for deployment without the presence of passport. Once IP infringement occurs, the owner can replace the passport-free layers of the suspected model with the corresponding passport-aware layers, and extract the signature from the signs of $wp_\gamma$ to prove the ownership.

\begin{figure}[t]
    \centering
    \begin{subfigure}{0.90\linewidth}
        \includegraphics[width=\textwidth]{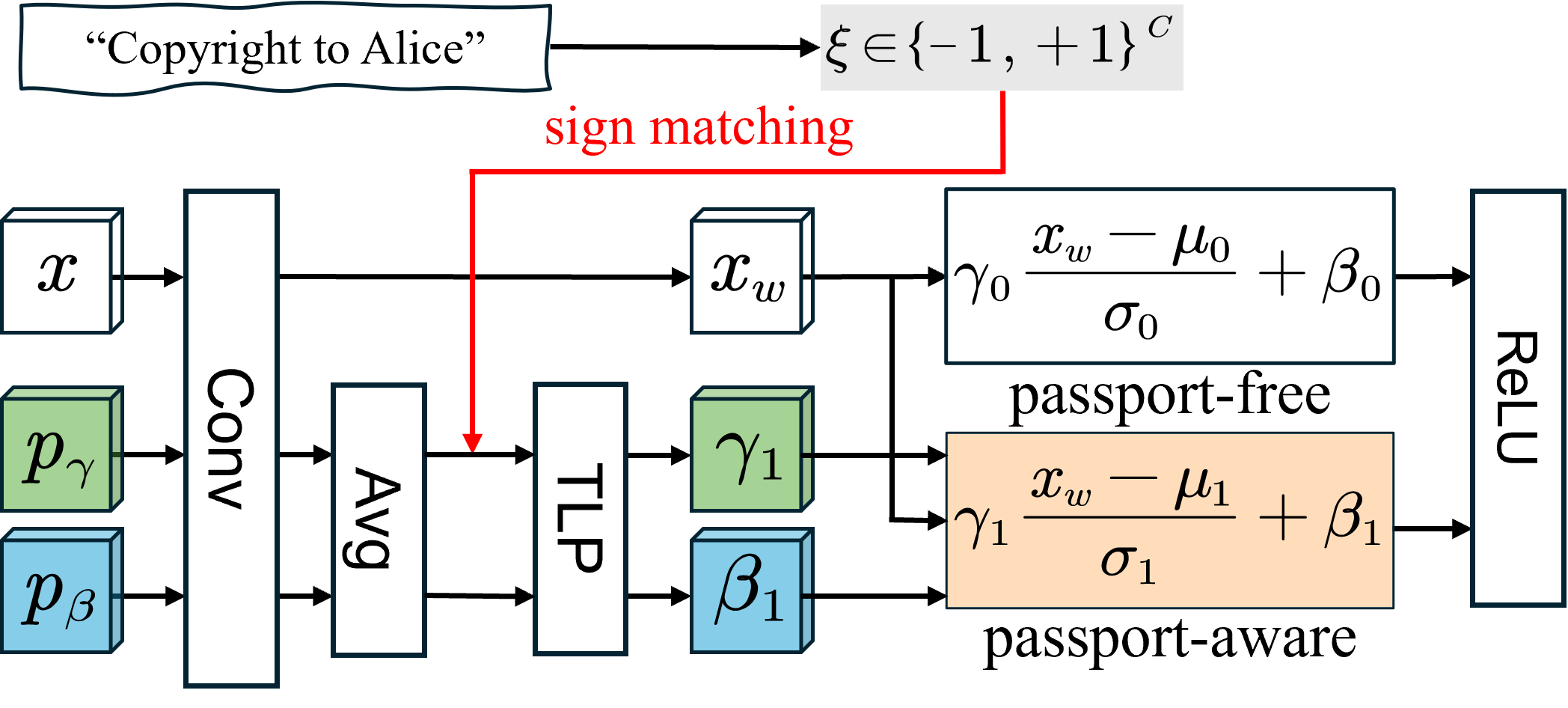}
        \caption{Passport layer of PAN. The passport-free and passport-aware layers utilize two groups of statistics mean/std.}
        \label{fig:PAN}
    \end{subfigure}
    \\
    \vspace{2mm}
    \begin{subfigure}{0.90\linewidth}
        \includegraphics[width=\textwidth]{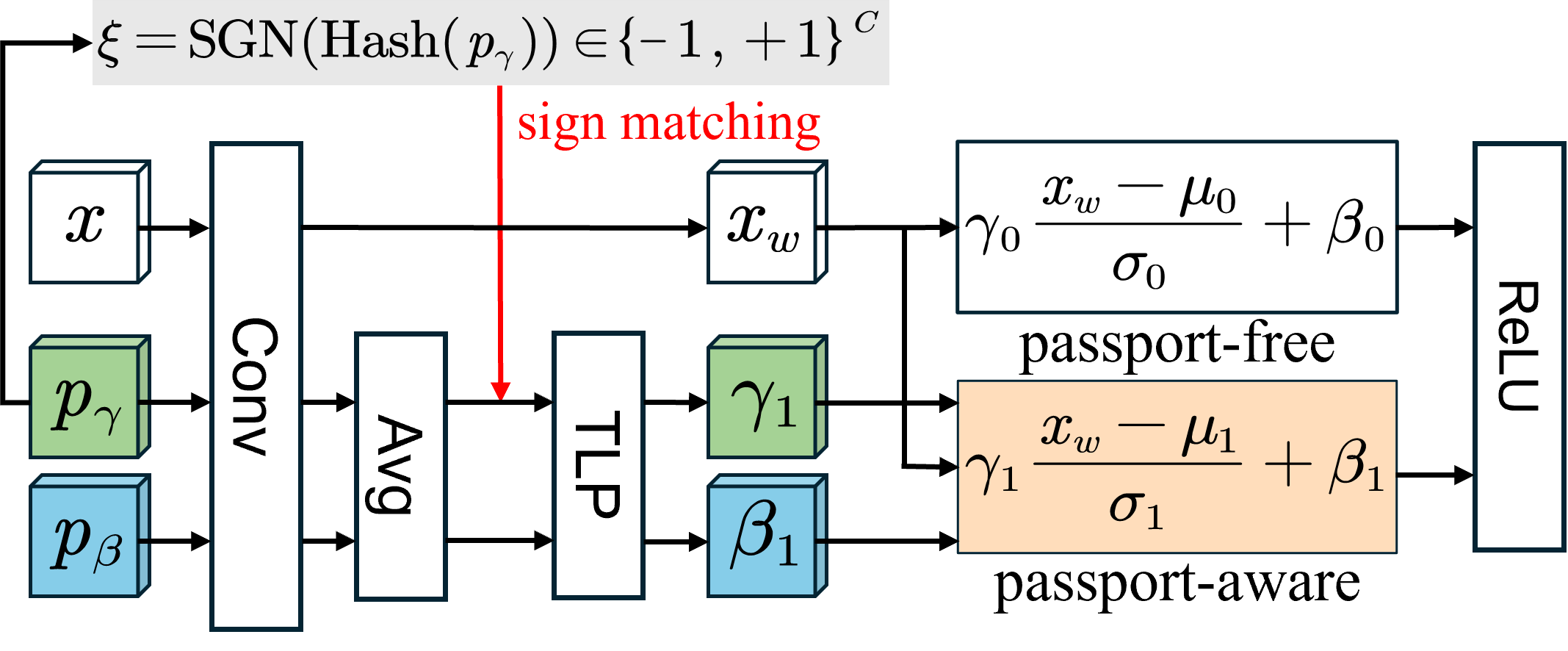}
        \caption{Passport layer of TdN. The passport $p_\gamma$ is first hashed to a $C$-bit binary sequence, and then mapped with $\operatorname{SGN}(\cdot)$ to generate the $\pm 1$ signature $\xi=\operatorname{SGN}(\operatorname{Hash}(p_\gamma))$.}
        \label{fig:TdN}
    \end{subfigure}

    \caption{Passport layers of (a) PAN and (b) TdN.}
    \vspace{-5mm}
\end{figure}

\subsubsection{Passport-Aware Normalization (PAN) ~\cite{zhang2020passport}}

Unfortunately, matching the signs of $wp_\gamma$ with $\xi$ is a strong constraint, which severely influences the feature extraction ability of the convolutional layer by backward propagation, and thus may lead to serious performance degradation on both branches. PAN addressed this issue with two improvements. As depicted in Fig.~\ref{fig:PAN}, PAN learns two groups of mean/std statistics separately for the two branches. To alleviate the sign matching constraint on the convolutional layer,
PAN also introduces an extra two-layer perceptron (TLP) after the average pooling layer to locally project the affine factors to proper scales as:
\begin{equation}
    \gamma_1 = \operatorname{TLP}(wp_\gamma), \quad \beta_1 = \operatorname{TLP}(wp_\beta).
\end{equation}

\subsubsection{Trapdoor Normalization (TdN) ~\cite{liu2023trapdoor}} \label{sec:TdN}

However, both DeepIPR and PAN remain vulnerable to ambiguity attacks with oracle passports. Due to the large parameter space of the passport, it is feasible for the adversary to generate forged passports that differ largely from the original one, while still matching the signature and retaining the target model's utility.

\textbf{Definition 1} (Ambiguity attacks with oracle passports~\cite{liu2023trapdoor}). \textit{Given a protected model, the original passport $p$, and the signature $\xi$, a forge passport $\tilde{p} = \{\tilde{p}_\gamma, \tilde{p}_\beta\}$ can be created within the feasible perturbation space $\delta(p)$ with respect to $p$ by solving the following bi-level optimization problem:
\begin{equation}
    \vspace{-1mm}
    \begin{aligned}
        \min   \quad &  \mathbb{E}_{(\mathbf{x}_i, \mathbf{y}_i) \sim D_\text{sub}} \big[\mathcal{L}_s + \lambda \cdot \mathcal{L}_a\big], \\
        {s.t.} \quad &  \tilde{p} = \underset{\tilde{p} \in \delta(p)}{\arg\max} \operatorname{Dis}(p, \tilde{p}),
    \end{aligned}
    \vspace{-1mm}
    \label{eq:oracle}
\end{equation}
where $D_{sub} \subset D_{tr}$ denotes a small subset of training data available to the attacker, and $\operatorname{Dis}(\cdot)$ measures the distance between $p$ and $\tilde{p}$. The forged $\tilde{p}$ is significantly different from the original $p$ but achieves comparable inference performance, while keeping $\xi$ and the protected model unchanged.}

TdN thwarts this attack by using a hash function of the genuine passport as the owner signature. As shown in Fig.~\ref{fig:TdN}, instead of being directly converted from a pre-defined text message, the $C$-bit signature $\xi=\operatorname{SGN}(\operatorname{Hash}(p_\gamma))$ is created by hashing the passport $p_\gamma$ with a pre-defined hash function, where $\operatorname{SGN}(\cdot): \{0, 1\}^n \rightarrow \{-1, +1\}^n$ denotes a mapping function that converts the binary hash result to a $\pm 1$ string. The one-way hash prevents the forged passport $\tilde{p}$ obtained by (\ref{eq:oracle}) from mapping to $\xi$ as
\begin{equation}
    \vspace{-1mm}
    \xi = \operatorname{SGN}(\text{Hash}(p_\gamma)) \neq \operatorname{SGN}(\text{Hash}(\tilde{p}_\gamma))
    \vspace{-1mm}
\end{equation}
holds except with negligible probability.

Since it is computationally intractable to reverse a hash function or generate a collision based on the hash value (signature), the attacker cannot create a counterfeit passport that passes the verification. One limitation of making the signature passport-dependent is the signature can no longer be freely designated and kept private from legitimate users or attackers who acquire the passport.

\subsubsection{Steganographic Passport (SteP) ~\cite{cui2024steganographic} }

Only until recently, both passive ownership proof and multi-user active control without retraining are achieved by Steganographic Passport (SteP)~\cite{cui2024steganographic}. All the aforementioned passport-based methods before SteP focus solely on overcoming ambiguity attacks by providing a stronger non-repudiable ownership claim upon model infringement.

Given a pre-trained invertible steganography network $\mathcal{S}(\cdot; \cdot)$, the copyright text $\mathcal{T}$, and the original passport images ($I = \{I_\gamma, I_\beta\}$), the owner passport images ($I_o = \{I_{o_\gamma}, I_{o_\beta}\}$) can be created as $I_{o_\gamma} = \mathcal{S}(I_\gamma; \mathcal{T})$ and $I_{o_\beta} =\mathcal{S}(I_\beta; \mathcal{T})$. $\mathcal{S}(\cdot;\cdot)$ imperceptibly embeds the copyright text $\mathcal{T}$ into owner passport images without introducing visible perturbations. Similar to TdN, SteP further trains the target model with the owner passport $p_o = \{p_{o_\gamma}, p_{o_\beta}\}$ derived from $I_o = \{I_{o_\gamma}, I_{o_\beta}\}$. Each end-user is provided with unique passport images $I_{u} = \{I_{u_\gamma}, I_{u_\beta}\}$, which appear visually identical to $I_o$ but contain an imperceptibly embedded user ID. In this context, the one-way correlation is only established from $I_{o_\gamma}$ to the signature, and the user passport images fail to prove the ownership due to the the avalanche effect of hash function.

SteP achieves active control in the online MLaaS scenario. Specifically, each end-user provides the cloud server with $I_{u}$ for authentication. The end-user is validated to be a legal user only if a recorded user ID can be successfully recovered from $I_{u}$. However, the active control of SteP is not applicable to the offline mode, because the owner has no control over the inference stage.

To fill the void in existing passport-based methods, we propose CHIP -- a more versatile and robust framework that (1) resists ambiguity attacks with oracle passports; (2) provides flexibility in certification of licensor and licensees without requiring extensive model retraining; and (3) offers active control on both online and offline modes.

\begin{figure*}[t]
    \centering
    \includegraphics[width=0.95\linewidth]{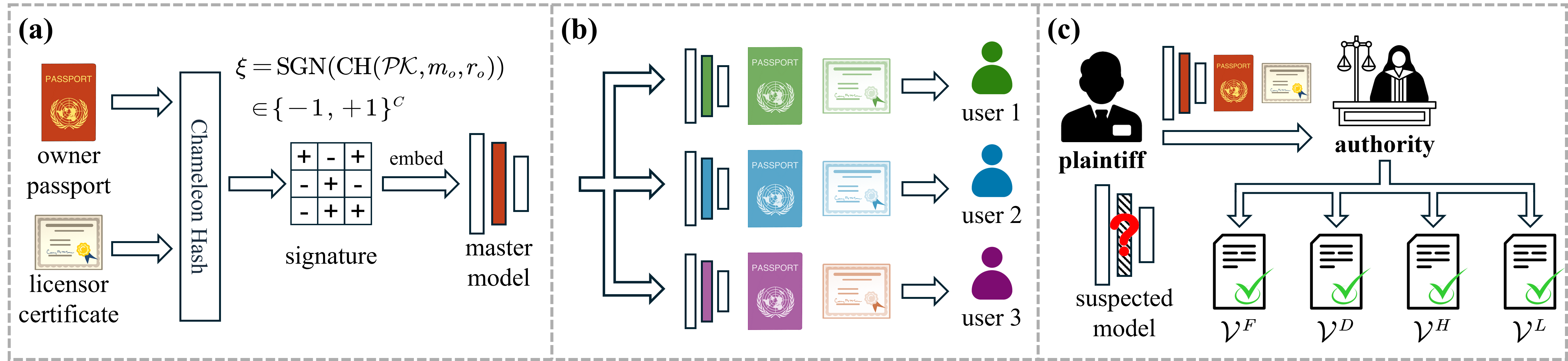}
    \caption{The proposed method contains three main stages: (a) Master model watermarking; (b) User triplet generation and distribution; (c) Ownership verification and traitor tracing.}
    \vspace{-4mm}
    \label{fig:pipeline}
\end{figure*}

\section{Chameleon Hash-based Irreversible Passport}

\subsection{Overview}

Table S2 of the Supplementary Material summarizes the key notations used throughout this paper. Fig.~\ref{fig:pipeline} depicts the overall pipeline of CHIP, which consists of three main stages:
\begin{description}
    \item[(a) Master model watermarking.] The owner initializes an owner passport $p_o = \{p_{o_\gamma}, p_{o_\beta}\}$ and a licensor certificate $r_o$ converted from the copyright text $\mathcal{T}$, and uses them to calculate an immutable signature $\xi$ with the chameleon hash function. The target model is then trained with $p_o$ and $\xi$ to obtain a watermarked master model $\mathcal{M}_o$.
    \item[(b) User triplet generation and distribution.] Instead of releasing $\mathcal{M}_o$, the owner creates $N$ unique triplets $\{\mathcal{M}^j_u, p^j_u, r^j_u\}^N_{j=1}$ from $\mathcal{M}_o$ by trapdoor collision. Each triplet is uniquely distributed to a registered user. A registered user $u_j$ of model $\mathcal{M}^j_u$ can use the model normally with its assigned passport $p^j_u=\{p^j_{u_\gamma}, p^j_{u_\beta}\}$, and prove his use permission upon request by presenting his unique licensee certificate $r^j_u$.
    \item[(c) Ownership verification and traitor tracing.] On model IP infringement, the owner presents $p_o$ and $r_o$ to validate the chameleon hash signature $\xi$ recovered from the suspected model. Additionally, the source of the model leakage and infringement can be traced by identifying the user passport.
\end{description}

\subsection{Master Model Watermarking} \label{sec:stage_1}

As discussed in Sec.~\ref{sec:TdN}, TdN~\cite{liu2023trapdoor} employs a hash function to create passport-dependent signature to resist ambiguity attacks with oracle passports. During training, the signature loss $\mathcal{L}_s$ ensures that the signs of $wp_\gamma$ match the hashed signature $\xi=\operatorname{SGN}(\operatorname{Hash}(p_\gamma))$, thereby constraining the convolutional kernel $\mathbf{W}$. However, by coupling usage control passport with ownership verification signature via a collision-resistant hash, the framework inherently limits the ability to achieve multi-user active control without retraining. Hence, to embedded $N$ distinct groups of passports and signatures into $N$ protected models, the owner must train the model from scratch for $N$ times. This approach becomes impractical as the number of users grows, posing a significant scalability challenge in real-world applications.

\begin{figure}[t]
    \centering
    \includegraphics[width=0.90\columnwidth]{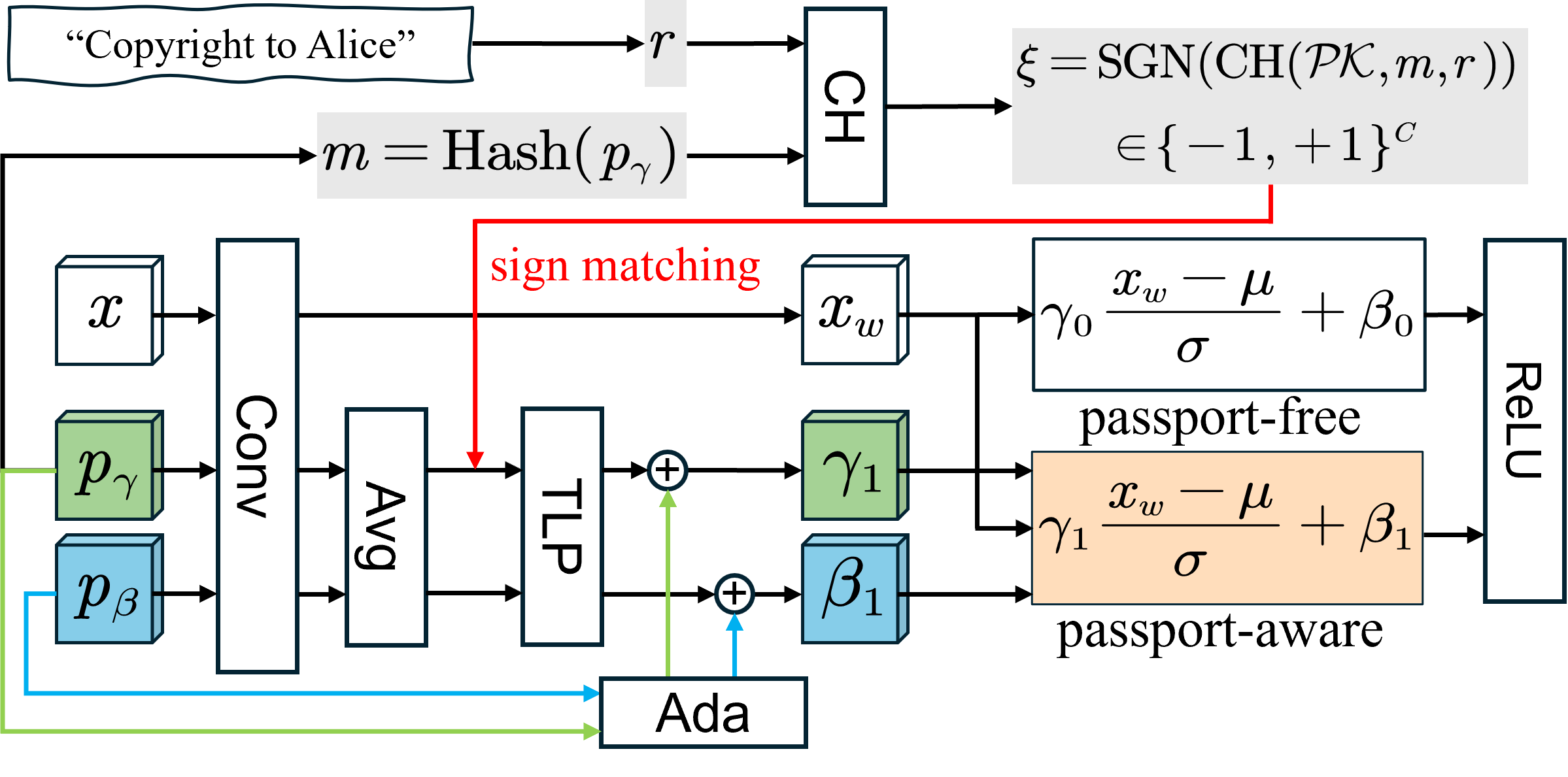}
    \caption{Passport layer of CHIP. The signature is created by chameleon hash, and we add a skip connection from the passport to passport-aware branch's affine factors (i.e., $\gamma_1$ and $\beta_1$) to enable effective active control.}
    \vspace{-3mm}
    \label{fig:CHIP}
\end{figure}

To solve the dilemma, CHIP creates the signature $\xi$ using a chameleon hash function. Fig.~\ref{fig:CHIP} presents the signature generation process and the structure of a CHIP layer. Given the owner passport $p_o = \{p_{o_\gamma}, p_{o_\beta}\}$, the copyright text $\mathcal{T}$, and the public key $\mathcal{PK}$ and the secret key $\mathcal{SK}$ defined by a chameleon hash function, an \textbf{immutable} signature can be generated for embedding as follows:
\begin{equation}
    \xi = \operatorname{SGN}(\operatorname{CH}(\mathcal{PK}, m_o, r_o)),
\end{equation}
where $m_o = \operatorname{Hash}(p_{o_\gamma})$ denotes a message digest derived from $p_{o_\gamma}$ with a standard hash function (e.g., SHA-512 in this work), and $r_o$ is an integer, referred to as a licensor certificate, which directly encodes $\mathcal{T}$.

The chameleon hash-based signature offers three important merits without conflicts: (1) Without knowledge of the secret key $\mathcal{SK}$, the mapping from $p_{o_\gamma}$ to $\xi$ remains irreversible, effectively addressing the weak resistance of~\cite{fan2019rethinking,fan2021deepipr,zhang2020passport} to ambiguity attacks with oracle passports; (2) The licensor certificate $r_o$ acts as a secure independent compactor of copyright information to overcome the restriction of TdN~\cite{liu2023trapdoor}, and enhance the credibility of ownership verification; (3) The trapdoor-collision property of chameleon hash enables the owner to generate diverse user-specific passports to support active control without retraining the model from scratch.

The owner passport and chameleon hash-based signature are further utilized to train the master model $\mathcal{M}_o$. As shown in Fig.~\ref{fig:CHIP}, the architecture of the CHIP layer is similar to that of the PAN or TdN layer but with two modifications.

First, we observe that the passport-aware branch exhibits a certain degree of tolerance to passport errors. As a result, $\mathcal{M}^\text{a}$ can still produce normal predictions even when presented with a forged passport that slightly deviates from the correct one. This behavior can be primarily attributed to the fact that the convolutional, pooling, and TLP layers tend to suppress subtle features in the passport, leading to the generation of similar affine factors. This limitation undermines the effectiveness of active control, as the target model's usage is not strictly bound to a unique passport. To address this issue, we introduce a skip connection that directly links the passport to the affine factors, which is formulated as follows:
\begin{equation}
    \begin{aligned}
        \gamma_1 = \operatorname{Ada}(p_{o_\gamma}) + \operatorname{TLP}(wp_{o_\gamma}),\\
        \beta_1 = \operatorname{Ada}(p_{o_\beta}) + \operatorname{TLP}(wp_{o_\beta}),
    \end{aligned}
\end{equation}
where $wp_{o_\gamma} = \operatorname{Avg}(\mathbf{W} \otimes p_{o_\gamma})$, $wp_{o_\beta} = \operatorname{Avg}(\mathbf{W} \otimes p_{o_\beta})$, and $\operatorname{Ada}(\cdot)$ denotes the adaptive pooling function used to downsample $p_{o_\gamma}$ and $p_{o_\beta}$ to match the dimensions of $\gamma_1$ and $\beta_1$. By incorporating this skip connection, the values of $\gamma_1$ and $\beta_1$ become highly dependent on $p_{o_\gamma}$ and $p_{o_\beta}$. Consequently, \textbf{a mismatched passport always results in significant performance degradation, thus achieving successful active control}.

Second, the two branches share the same set of mean and std statistics. To ensure consistency between their affine factors, we introduce a balance loss defined as:
\begin{equation}
    \mathcal{L}_\text{bal} = \ell_1(\gamma_0, \gamma_1) + \ell_1(\beta_0, \beta_1),
    \label{eq:balance_loss}
\end{equation}
where $\ell_1(\cdot,\cdot)$ measures the $\ell_1$ loss. This design aims to minimize the performance deviation between the two branches: when their affine factors are close, they are more likely to produce highly similar outputs.

The master model is trained and watermarked through the following joint optimization objective:
\begin{equation}
    \mathcal{L}_\text{total} = \mathcal{L}_\text{f} + \mathcal{L}_\text{a} + \mathcal{L}_\text{s} + \mathcal{L}_\text{bal}.
    \label{eq:overall_loss}
\end{equation}

The pseudo-code for the master model watermarking process is shown in Algorithm S3 of the Supplementary Material.

\subsection{User Triplet Generation and Distribution} \label{sec:stage_2}

To incorporate active control and licensee tracing into distributed passport-aware models, we turn to \textbf{Definition 1} to reverse the malevolent passport forging attack into a benevolent generator of user passports that can be tied to the signature $\xi$ by trapdoor collision of chameleon hash.

\textbf{Definition 2} (User triplet generation). \textit{For each registered user $u_j$, the owner optimizes the TLP layer and a copy of owner passport $p_o$ to obtain a unique triplet $\{\mathcal{M}^j_u, p^j_u, r^j_u\}$ with the following bi-level objective:
\begin{equation}
    \begin{aligned}
        \min   \ &  \left[\mathcal{L}_\text{s} + \mathcal{L}_\text{bal}\right], \\
        {s.t.} \ &  p^j_u = \underset{p^j_u \in \delta(p_o)} {\arg\max} \left[\operatorname{Dis}(p_o, p^j_u) + \sum_{k=1}^{j-1}\operatorname{Dis}(p^k_u, p^j_u)\right].
    \end{aligned}
    \label{eq:self_ambiguity}
\end{equation}
A unique licensee certificate for $u_j$ can then be generated by trapdoor collision as $r^j_u = \operatorname{Col}(\mathcal{SK}, m_o, r_o, m^j_u)$, where $m^j_u = \operatorname{Hash}(p^j_{u_\gamma})$.}

In the first line of Eq.~\eqref{eq:self_ambiguity}, $\mathcal{L}_\text{s}$ ensures persistent signature embedding to the user model $\mathcal{M}^j_u$, while $\mathcal{L}_\text{bal}$ preserves the user model performance in the presence of its designated user passport $p^{j}_u$. The second line of Eq.~\eqref{eq:self_ambiguity} forces $p^j_u$ to be different from the owner passport $p_o$ and previously generated user passports, i.e., $\{p^1_u, p^2_u, \cdots, p^{j-1}_u\}$. This is a data-free optimization, which makes the user triplet generation process flexible and efficient.

The trapdoor collision guarantees that $\operatorname{CH}(\mathcal{PK}, m_o, r_o) = \operatorname{CH}(\mathcal{PK}, m^j_u, r^j_u)$ holds $\forall j \in \{1, 2, \cdots, N\}$. In other words, all user passports can be successfully mapped to the immutable signature $\xi$ without re-watermarking as
\begin{equation}
    \begin{aligned}
        \xi & = \operatorname{SGN}(\operatorname{CH}(\mathcal{PK}, m_o, r_o)) \\
            & = \operatorname{SGN}(\operatorname{CH}(\mathcal{PK}, m^j_u, r^j_u)).
    \end{aligned}
\label{eq:CH_match}
\end{equation}
Since all licensee certificates are generated by trapdoor collision, no meaningful text string can be decoded from them. Therefore, \textbf{the ownership information is only plaintext encoded in the licensor certificate $r_o$}.

Unlike ambiguity attacks with oracle passports (\textbf{Definition 1}), we not only optimize the user passport but also fine-tune the TLP layer of the passport-aware branch locally, such that each user model $\mathcal{M}^j_u$ is bound exclusively to its user passport $p^j_u$. Applying a user passport $p^k_u, k \neq j$ from a different user model to $\mathcal{M}^j_u$ will result in significant performance degradation. As the parameter amount of the TLP and the passport is small, the computational cost of producing a user model is considerably lower than retraining or re-watermarking the model, making the triplet generation for individual users highly efficient. The owner keeps the master model private, and sells to each registered user $u_j$ a unique distributed passport-aware user model $\mathcal{M}^j_u$ with its exclusive passport $p^j_u$ and licensee certificate $r^j_u$.

Algorithm S4 of the Supplementary Material delineates the process of creating multiple user triplets.

\subsection{Verification and Tracing} \label{sec:stage_3}

Let $\dot{\mathcal{M}}$ denote a suspected model. The verification stage involves three parties: the \textit{plaintiff} who claims the ownership of $\dot{\mathcal{M}}$; the \textit{defendant} who is accused of model infringement or abuse; and an \textit{authority} (e.g., a copyright tribunal or court) with jurisdiction in dispute resolution.

The $\textit{plaintiff}$ presents the passport $p$, the certificate $r$, the signature $\xi$, and passport-aware branch to the $\textit{authority}$. The \textit{authority} replaces the norm layer of $\dot{\mathcal{M}}$ with the provided passport-aware branch and conduct the following four tests.

The \textbf{performance fidelity test} $\mathcal{V}^F$:
\begin{equation}
    \mathcal{V}^F \Longleftrightarrow \mathbb{E}_{(\mathbf{x}_i, \mathbf{y}_i) \sim D_\text{ts}} \left\{\mathbb{I}\big[\dot{\mathcal{M}}(\mathbf{x}_i), \mathbf{y}_i\big]\right\} > \tau_\text{fidelity},
\end{equation}
where $\tau_\text{fidelity}$ is the minimal inference accuracy threshold, and $D_{ts}$ denotes the test dataset. $\mathbb{I}\big[\cdot, \cdot\big]$ is an indicator function which returns 1 if the two inputs are the same, and 0 otherwise. $\mathcal{V}^F$ evaluates the fidelity of $\dot{\mathcal{M}}$. Passing $\mathcal{V}^F$ indicates that the passport $p$ submitted by the $\textit{plaintiff}$ can operate the suspected model normally with the passport-aware branch.

The \textbf{signature detection test} $\mathcal{V}^D$:
\begin{equation}
    \mathcal{V}^D \Longleftrightarrow \psi = \frac{1}{C} \sum^{C}_{i=1} (\xi^\ast \wedge \xi) > 1 - \tau_\text{error},
\end{equation}
where $\psi$ is the signature detection accuracy (SDA) which measures the proportion of matching bits between the extracted signature $\xi^\ast = \operatorname{sign}(wp_\gamma)$ and the provided signature $\xi$. $\tau_\text{error}$ is a is a pre-defined small error tolerance (5\%) between these two signatures. A high SDA validates that the passport can be correctly convoluted to a designated signature.

The \textbf{passport hashing test} $\mathcal{V}^H$:
\begin{equation}
    \mathcal{V}^H \Longleftrightarrow \phi = \frac{1}{C} \sum^{C}_{i=1} (\xi^\ast \land \xi^\prime) > 1 - \tau_\text{error},
\end{equation}
where $\xi^\prime = \operatorname{SGN}(\operatorname{CH}(\mathcal{PK}, \operatorname{Hash}(p_\gamma), r))$ denotes the signature computed by the chameleon hash using $p$ and $r$. The passport hashing accuracy (PHA), denoted by $\phi$, measures the proportion of matching bits between $\xi^\ast$ and $\xi^\prime$. Passing $\mathcal{V}^H$ verifies the ``chameleon'' signature generated by $p$ and $r$ can produce the intended collision with the extracted signature.

Lastly, the \textbf{licensor test} $\mathcal{V}^L$:
\begin{equation}
    \mathcal{V}^D \Longleftrightarrow \operatorname{Dec}(r),
\end{equation}
where $\operatorname{Dec}(\cdot)$ is the ASCII decoding operation. $\mathcal{V}^L$ confirms the validity of the ownership claim. Only the owner licensor certificate can be decoded to a legible and meaningful copyright text. All licensee certificates are random hashed values that cannot be decoded to meaningful texts.

Note that registered users can also pass $\mathcal{V}^F$, $\mathcal{V}^D$, and $\mathcal{V}^H$, but not $\mathcal{V}^L$. Only when all four tests are passed can an ownership claim be confirmed. Once the ownership is validated, the culprit responsible for the model infringement can be traced by subjecting user passports in the plaintiff's repository to the fidelity test in turn. The registered user whose passport passes $\mathcal{V}^F$ is identified as the culprit. Table S3 of the Supplementary Material summarizes the goals of the four tests.

\begin{table*}[t]
    \centering
    \caption{Inference accuracy (\%) of passport-free/passport-aware models across four datasets and two architectures. The first row “clean” represents unprotected models without passport layers. ``+bd'' denotes the combination of the passport-based method with a backdoor watermark~\cite{uchida2017embedding}. Both ``CHIP+bd'' and ``CHIP'' are evaluated on watermarked master models. The highest average accuracy for passport-free/passport-aware models is highlighted in bold in the last column.}
    
    \begin{minipage}{\textwidth}
        \centering
        \renewcommand\arraystretch{1.1}
        \resizebox{1.0\linewidth}{!}{
            \begin{tabular}{l|cc|cc|cc|cc|c}
                \Xhline{2\arrayrulewidth}
                \multicolumn{1}{c|}{\multirow{2}{*}{AlexNet}}& \multicolumn{2}{c|}{CIFAR-10} & \multicolumn{2}{c|}{CIFAR-100}          & \multicolumn{2}{c|}{Caltech-101}        & \multicolumn{2}{c|}{Caltech-256}                             & \multirow{2}{*}{Mean}          \\ \cline{2-9}
                \multicolumn{1}{c|}{}                        & BN            & GN            & \multicolumn{1}{c|}{BN} & GN            & \multicolumn{1}{c|}{BN} & GN            & \multicolumn{1}{c|}{BN} & GN                                 &                                \\ \Xhline{2\arrayrulewidth}
                \multicolumn{1}{l|}{clean}                   & 91.09         & 89.92         & 68.79                   & 65.05         & 72.20                   & 69.21         & 44.15                   & \multicolumn{1}{c|}{41.88}         & 67.79                          \\
                \multicolumn{1}{l|}{DeepIPR}                 & 86.17 / 89.50 & 89.06 / 88.34 & 32.70 / 64.04           & 62.80 / 60.79 & 65.59 / 64.29           & 66.89 / 66.72 & 38.28 / 39.89           & \multicolumn{1}{c|}{40.34 / 35.02} & 60.23 / 63.57                  \\
                \multicolumn{1}{l|}{PAN}                     & 91.12 / 90.87 & 89.89 / 89.47 & 68.14 / 68.09           & 64.50 / 63.38 & 71.81 / 71.27           & 68.59 / 66.21 & 44.72 / 41.25           & \multicolumn{1}{c|}{41.18 / 39.68} & 67.49 / 66.28                  \\
                \multicolumn{1}{l|}{TdN}                     & 91.27 / 91.37 & 90.12 / 89.80 & 68.14 / 67.57           & 64.67 / 63.79 & 70.90 / 68.64           & 67.80 / 66.67 & 43.96 / 42.32           & \multicolumn{1}{c|}{41.36 / 39.37} & 67.28 / 66.19                  \\
                \multicolumn{1}{l|}{SteP}                    & 91.62 / 91.63 & 89.92 / 89.72 & 68.02 / 67.28           & 64.91 / 61.95 & 71.19 / 70.11           & 69.89 / 67.91 & 44.20 / 41.99           & \multicolumn{1}{c|}{41.84 / 38.59} & \textbf{67.70} / 66.15         \\
                \multicolumn{1}{l|}{CHIP+bd (Ours)}          & 90.70 / 90.73 & 89.44 / 89.48 & 68.57 / 68.58           & 64.91 / 64.92 & 71.53 / 71.53           & 68.70 / 68.64 & 44.29 / 44.27           & \multicolumn{1}{c|}{40.49 / 40.47} & 67.33 / 67.33                  \\
                \multicolumn{1}{l|}{CHIP (Ours)}             & 91.45 / 91.48 & 90.07 / 90.05 & 68.77 / 68.78           & 64.37 / 64.38 & 71.69 / 71.69           & 68.93 / 68.93 & 44.80 / 44.82           & \multicolumn{1}{c|}{41.00 / 40.98} & 67.64 / \textbf{67.64}         \\ \Xhline{2\arrayrulewidth}
            \end{tabular}
        }
    \end{minipage}
    
    \vspace{2mm}
    
    \begin{minipage}{\textwidth}
        \centering
        \renewcommand\arraystretch{1.1}
        \resizebox{1.0\linewidth}{!}{
            \begin{tabular}{l|cc|cc|cc|cc|c}
                \Xhline{2\arrayrulewidth}
                \multicolumn{1}{c|}{\multirow{2}{*}{ResNet-18}}& \multicolumn{2}{c|}{CIFAR-10} & \multicolumn{2}{c|}{CIFAR-100}          & \multicolumn{2}{c|}{Caltech-101}        & \multicolumn{2}{c|}{Caltech-256}                             & \multirow{2}{*}{Mean}                   \\ \cline{2-9}
                \multicolumn{1}{c|}{}                          & BN            & GN            & \multicolumn{1}{c|}{BN} & GN            & \multicolumn{1}{c|}{BN} & GN            & \multicolumn{1}{c|}{BN} & GN                                 &                                         \\ \Xhline{2\arrayrulewidth}
                \multicolumn{1}{l|}{clean}                     & 95.00         & 93.48         & 76.39                   & 72.16         & 70.68                   & 66.67         & 53.73                   & \multicolumn{1}{c|}{45.38}         & 71.69                                   \\
                \multicolumn{1}{l|}{DeepIPR}                   & 93.17 / 92.89 & 90.52 / 90.56 & 67.35 / 71.54           & 68.19 / 67.76 & 65.37 / 67.29           & 60.11 / 59.66 & 41.50 / 45.46           & \multicolumn{1}{c|}{42.45 / 41.35} & 66.08 / 67.06                           \\
                \multicolumn{1}{l|}{PAN}                       & 94.62 / 94.56 & 93.50 / 93.65 & 76.47 / 76.58           & 71.05 / 71.46 & 72.09 / 71.69           & 67.12 / 67.01 & 55.12 / 54.71           & \multicolumn{1}{c|}{44.70 / 43.94} & 71.83 / 71.70                           \\
                \multicolumn{1}{l|}{TdN}                       & 94.59 / 94.54 & 93.51 / 93.40 & 75.46 / 74.11           & 70.82 / 71.09 & 73.01 / 72.94           & 66.55 / 66.05 & 54.79 / 54.81           & \multicolumn{1}{c|}{44.65 / 44.19} & 71.67 / 71.39                           \\
                \multicolumn{1}{l|}{SteP}                      & 94.65 / 94.55 & 93.29 / 93.42 & 75.66 / 74.62           & 71.35 / 71.95 & 74.18 / 73.90           & 66.33 / 66.27 & 54.54 / 54.46           & \multicolumn{1}{c|}{43.43 / 43.66} & 71.68 / 71.60                           \\
                \multicolumn{1}{l|}{CHIP+bd (Ours)}            & 94.51 / 94.51 & 93.57 / 93.58 & 76.80 / 76.81           & 71.19 / 71.19 & 72.82 / 72.82           & 66.05 / 66.05 & 55.35 / 55.32           & \multicolumn{1}{c|}{45.28 / 45.24} & 71.95 / 71.94                           \\
                \multicolumn{1}{l|}{CHIP (Ours)}               & 94.80 / 94.79 & 93.51 / 93.51 & 76.64 / 76.64           & 70.91 / 70.91 & 72.54 / 72.60           & 67.74 / 67.68 & 55.04 / 55.07           & \multicolumn{1}{c|}{44.90 / 44.93} & \textbf{72.01} / \textbf{72.02}         \\ \Xhline{2\arrayrulewidth}
            \end{tabular}
        }
    \end{minipage}
    \vspace{-4mm}
    \label{tab:dep_ver_acc}
    
\end{table*}

\subsection{Cloud application} \label{sec:cloud_application}

The above CHIP usage control and protection mechanism applies to offline distributed models, which are safeguarded against legal buyers who have white-box access to their purchased models' architectures and weights. For models deployed as MLaaS, it is impractical to provide each registered user with a separate model for usage control.

Instead of creating $N$ user triplets, the owner simply generates $N$ distinct random passports $\{p^1_u, p^2_u, \cdots, p^{N}_u\}$ and compute the corresponding licensee certificates $\{r^1_u, r^2_u, \cdots, r^{N}_u\}$ by trapdoor collision. Each licensed user $u_j$ is issued a unique tuple $\{p^j_u, r^j_u\}$ that serves as the identity token. The API call from a user is approved if the provided tuple can generate a intended signature, i.e., $\operatorname{SGN}(\operatorname{CH}(\mathcal{PK}, m^j_u, r^j_u)) = \xi$. Subsequently, the master model can process user-provided test images for classification. In the cloud scenario where the model owner maintains full control over inference, we recommend using the passport-free branch for prediction. As demonstrated in our complexity analysis (Sec.~\ref{sec:further_analysis}), this branch offers superior computational efficiency in inference compared to the passport-aware branch.

CHIP provides great flexibility to the owner in the MLaaS scenario: revoking an expired passport or issuing a new passport and licensee certificate can be achieved at any time, through efficient chameleon hash computing.

Overall, since the cloud model is a black-box to users and the owner has full control over the inference stage, it turns out that with CHIP, single-model multi-user active control becomes simpler to implement on cloud. Hence, our evaluations mainly focus on the more challenging offline scenario.

\section{Evaluations}

\subsection{Experimental Settings}

\textbf{Baselines.} To benchmark CHIP, we compare it against four state-of-the-art passport-based methods: DeepIPR~\cite{fan2021deepipr}, PAN~\cite{zhang2020passport}, TdN~\cite{liu2023trapdoor}, and SteP~\cite{cui2024steganographic}. These methods are re-implemented using their official codes on GitHub.
% \footnote{DeepIPR: https://github.com/kamwoh/DeepIPR.}\footnote{PAN: https://github.com/ZJZAC/Passport-aware-Normalization.}\footnote{SteP: https://github.com/TracyCuiq/Steganographic-Passport.}

\textbf{Datasets and Networks.} To be consistent with the baselines, experiments are conducted on four image classification benchmarks, including CIFAR-10~\cite{krizhevsky2009learning}, CIFAR-100~\cite{krizhevsky2009learning}, Caltech-101~\cite{FeiFei2004LearningGV}, and Caltech-256~\cite{FeiFei2004LearningGV}. AlexNet~\cite{krizhevsky2012imagenet} and ResNet-18~\cite{he2016deep} architectures are used for the evaluation, with both BN~\cite{ioffe2015batch} and GN~\cite{wu2018group}.

\textbf{Implementation details.} All models are trained from scratch for 200 epochs, starting with an initial learning rate of 0.01, which decays by a factor of 0.1 at epochs 100 and 150. The SGD optimizer is used for training, with a weight decay of 5e-4. Following the configurations in~\cite{fan2019rethinking,fan2021deepipr}, the last three norm layers of AlexNet, and the norm layers within ``layer4'' of ResNet-18, are selected for passport embedding. We also conduct additional experiments on CHIP with backdoor watermarking~\cite{adi2018turning} to demonstrate their compatibility. For user triplet generation, the master model and the owner passport are optimized with 5,000 iterations, with a fixed learning rate of 0.01. Experiments are run on four NVIDIA A100 GPUs using Python 3.10.16 and PyTorch 2.6.0.

\subsection{Effectiveness and Verification Assessment}

Table~\ref{tab:dep_ver_acc} presents the inference performance of passport-free and passport-aware models. The first row ``clean'' in each sub-table shows the accuracy of unprotected models. The DeepIPR has lower accuracy for passport-free and passport-aware models since the convolutional layer's feature extraction ability is severely undermined by the signature embedding constraint (i.e., $\mathcal{L}_\text{s}$). In contrast, CHIP has comparable accuracy as PAN, TdN, and SteP that inserts a TLP layer after the pooling layer to alleviate the training difficulty, and surpasses them in some cases. Specifically, CHIP has the highest mean accuracy for both passport-free and passport-aware models on ResNet-18, and passport-aware model on AlexNet. For passport-free model on AlexNet, CHIP achieves the second highest mean accuracy of 67.64\%, which which is merely 0.06\% lower than that of SteP. These results indicate that CHIP's passport layers will not compromise the model's primary performance tasks.

Previous works witness an unignorable performance deviation between the passport-free and passport-aware branches. In general, the passport-aware model's performance is lower than that of the passport-free counterpart. For example, in the case of ``AlexNet\_Caltech-256\_BN\_PAN'', the passport-aware model is 3.47\% lower than that of the passport-free model. This level of performance degradation undermines the passport-aware models' utility for deployment. The balance loss $\mathcal{L}_\text{bal}$ of CHIP forces the affine factors of the two branches to be close, making their performance deviation extremely low.

The penultimate row in each sub-table presents the accuracies of the models that are jointly protected by CHIP and backdoor watermarking~\cite{uchida2017embedding}. The backdoor training samples cause only a slight performance drop in the passport-free and passport-aware models, demonstrating that CHIP can be easily supplemented by backdoor watermarking to provide additional means of infringement detection.

All methods in comparison can achieve 100\% SDA, i.e., the signature bits extracted from the passport layers perfectly match the actual embedded signature bits.

\begin{figure}[h]
    \centering
    \begin{subfigure}{0.45\linewidth}
        \includegraphics[width=\textwidth]{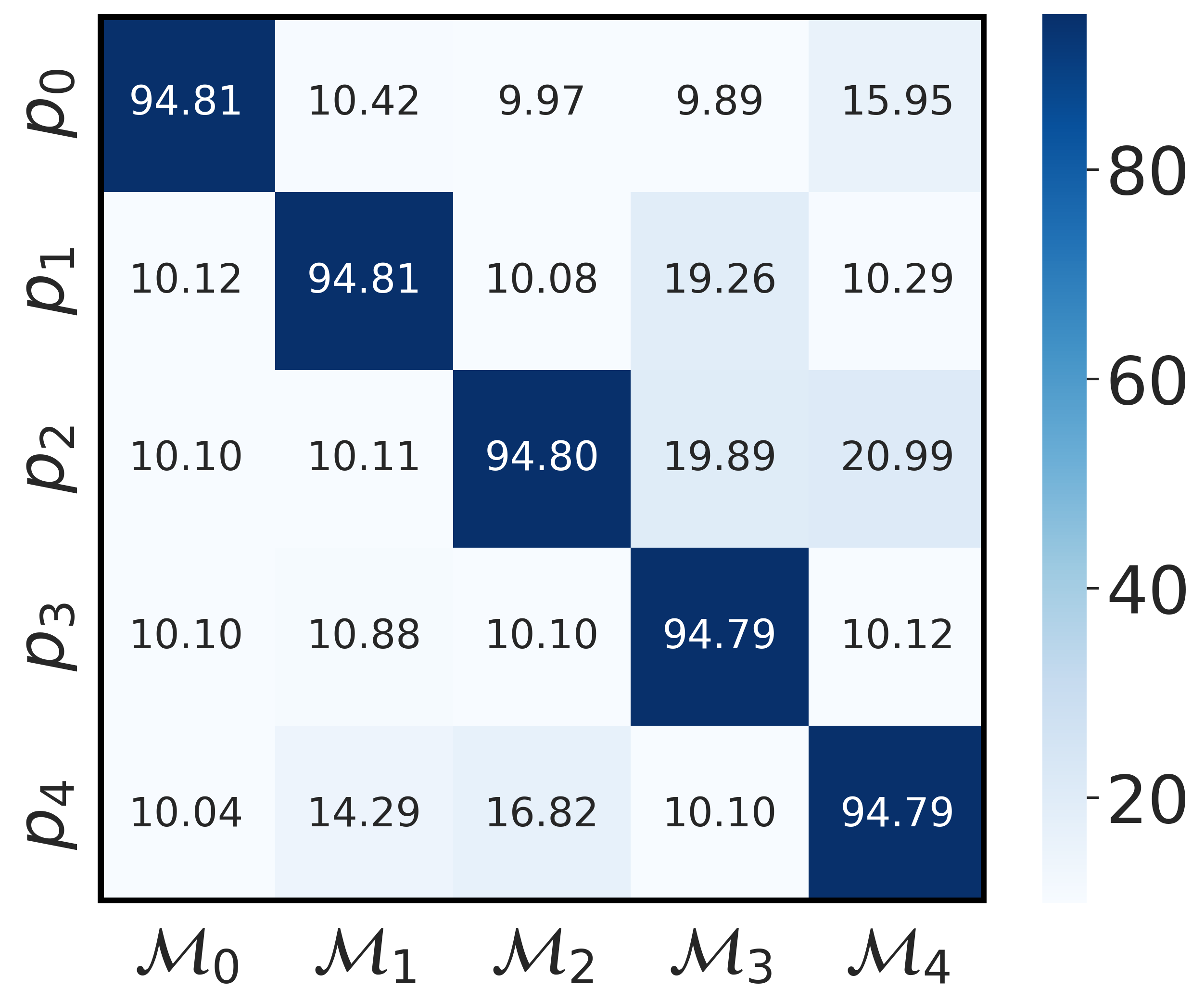}
        % \caption{Inference accuracy (\%)}
        \caption{}
        \label{fig:active_control_acc}
    \end{subfigure}
    \begin{subfigure}{0.45\linewidth}
        \includegraphics[width=\textwidth]{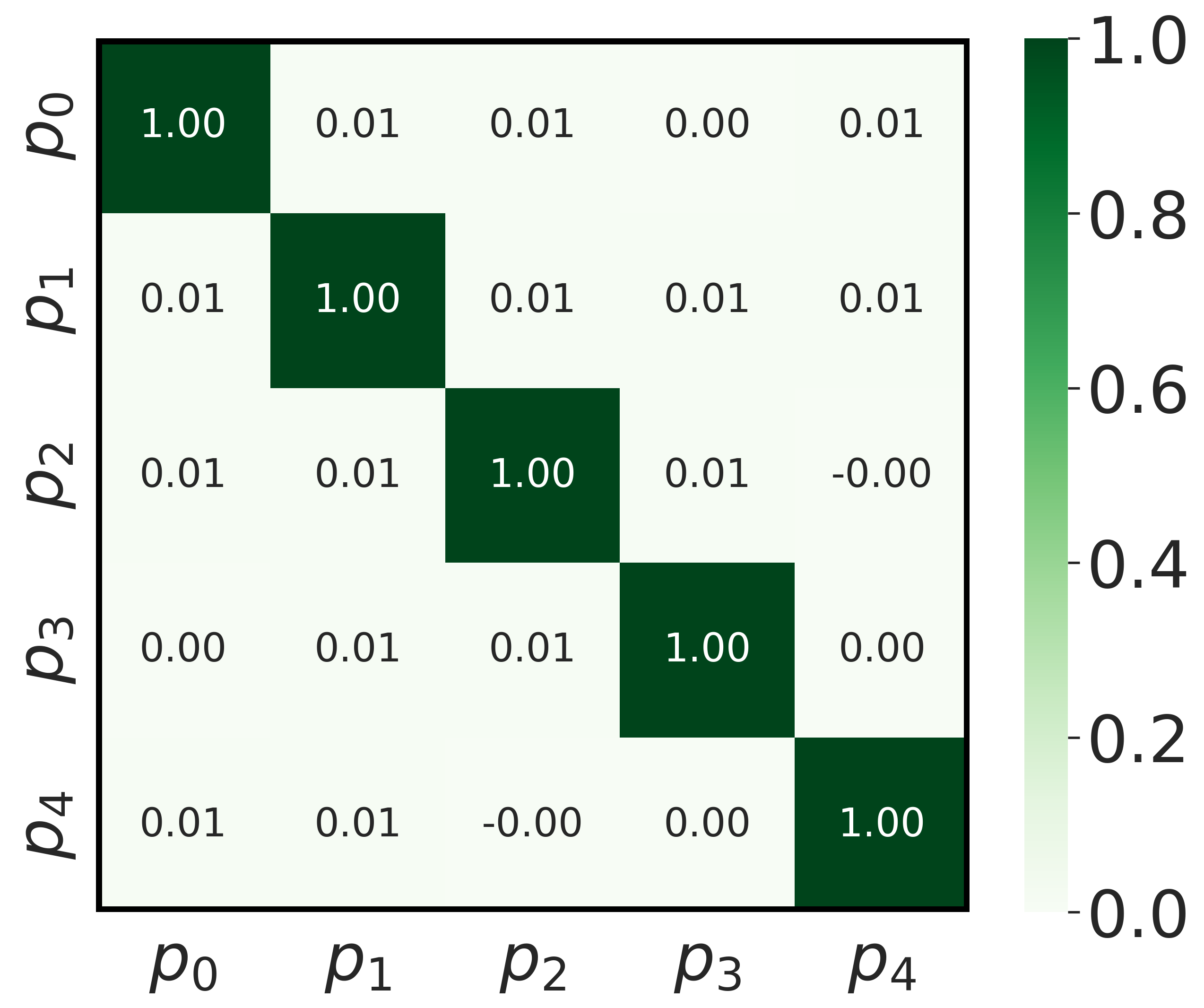}
        % \caption{Cosine similarity}
        \caption{}
        \label{fig:active_control_cs}
    \end{subfigure}

    \caption{(a) Confusion matrix for the performance evaluation of five different user passports on five user models. The inference accuracy of the unprotected model is 95.00\%. (b) Cosine similarity between user passports. Experiments are conducted on ``ResNet-18\_CIFAR-10\_BN''.}
    \vspace{-5mm}

\end{figure}

\subsection{Active Control Assessment}

We generate five user triplets based on the master model to test the active usage control. Taking the case of ``ResNet-18\_CIFAR-10\_BN'' as an example, Fig.~\ref{fig:active_control_acc} demonstrates that all five user models achieve high inference accuracy comparable to that of the clean model when the matched passport is presented. However, the inference accuracy drops dramatically to below 21\% with a mismatched passport, even if it is also a legal user passport issued by the owner. This highlights a significant advantage of CHIP over SteP~\cite{cui2024steganographic}, which fails to bind a distributed model to a unique passport. Hence, a user model can only be normally used by the unique user who holds the designated paired passport. As depicted in Fig.~\ref{fig:active_control_cs}, the cosine similarity between two distinct user passports is extremely small, indicating a high dissimilarity between them. This makes it easy to distinguish different user passports and conduct liable buyer traceability in the event of IP infringement. Furthermore, all user triplets achieve high SDA and PHA scores exceeding 99\%, confirming that both $\mathcal{V}^D$ and $\mathcal{V}^H$ can be successfully validated when the user model, user passport, and licensee certificate are properly matched.

Supplementary experimental results for active control assessment are provided in Fig. S1 and Fig. S2 of the Supplementary Material. These results conclude that CHIP achieves successful active control across all cases.

\subsection{Robustness}

The protected master model remains private to the owner. We assume that the attacker possesses a stolen user model, and in the worst-case scenario, even holds the corresponding user passport and the licensee certificate. The attacker may launch either ambiguity attacks to falsely claim ownership of the user model or removal attacks aimed at erasing the embedded watermark.

\noindent\textbf{Robustness against Ambiguity Attacks}

Two types of ambiguity attacks are considered: (1) Random passport attack. The attacker has no access to the correct passport and thus uses random passports to operate the stolen user model. (2) Ambiguity attack with oracle passport. As defined in \textbf{Definition 1}, without sacrificing the inference accuracy, the attacker creates a forged passport that can be projected to the original signature or a designated signature (e.g., flipping 10\% bits of the original signature).

\begin{figure}[t]
    \centering
    \includegraphics[width=1.0\columnwidth]{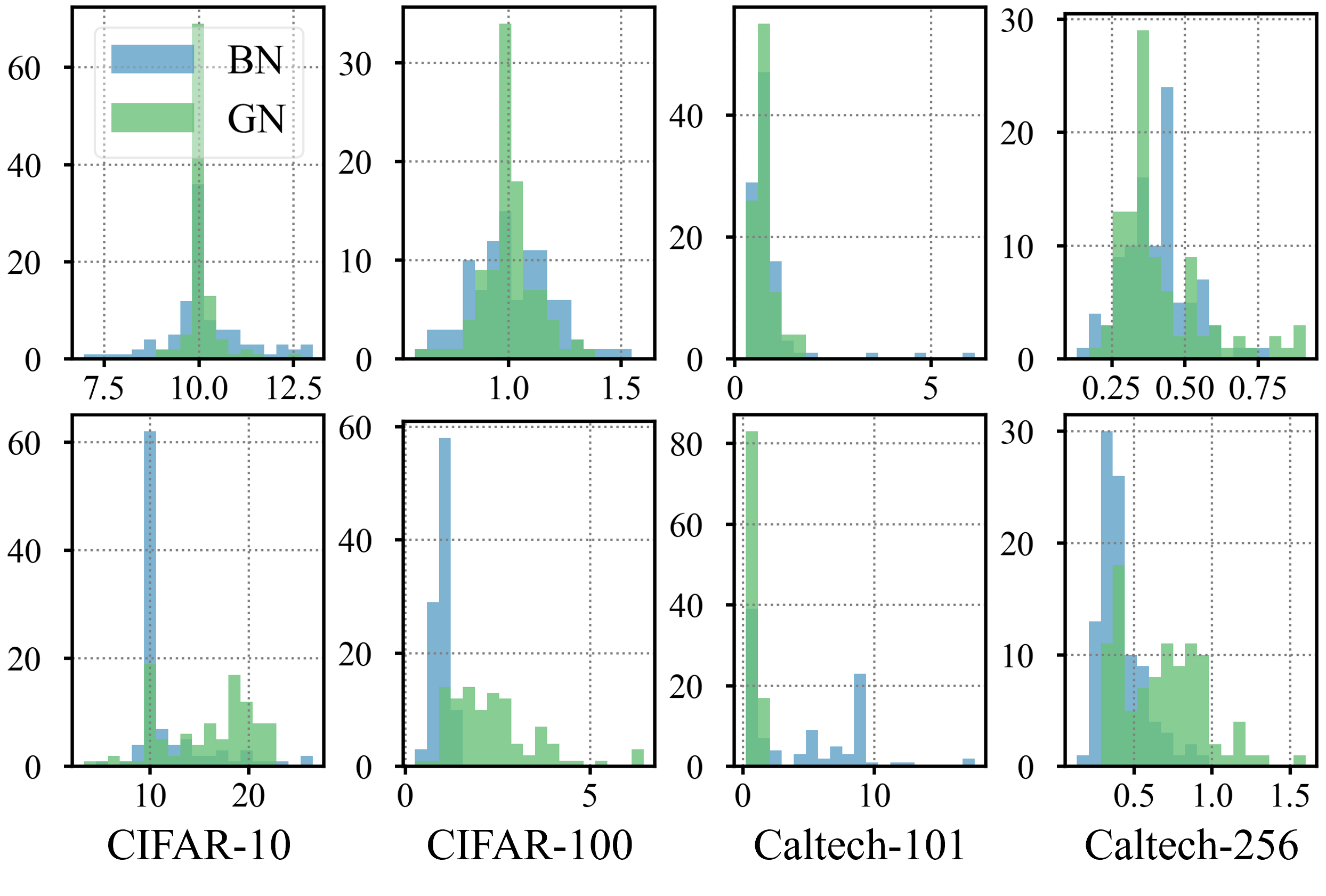}
    \caption{Performance of protected models under random passport attacks. In each histogram, the x-axis represents inference accuracy (\%), and the y-axis indicates frequency. The first row displays results for AlexNet, while the second row shows those for ResNet-18. Histograms for BN and GN are displayed in \textcolor[HTML]{5ea0c7}{blue} and \textcolor[HTML]{6bc179}{green}, respectively.}
    \vspace{-5mm}
    \label{fig:random_attack}
\end{figure}

\textit{Random passport attack.} To assess the robustness of each CHIP model against this attack, we generate 100 random passports and measure their inference accuracy on the stolen model. Fig.~\ref{fig:random_attack} presents the resulting histograms for each case over 100 test runs. Across all cases, the stolen model consistently displays an extremely low inference accuracy with distinct random passports. For example, with ``AlexNet\_CIFAR-10\_BN'', the 100 random passports yield an average accuracy of just 10.15\% ($\pm$1.07\%). Such a significantly degraded performance is comparable to that of an untrained model making random guesses. Consequently, false ownership claims using forged random passports fail the performance fidelity test $\mathcal{V}^F$. More critically, the attacker cannot even use the stolen model normally with forged passports.

\begin{table*}[t]
    \caption{Robustness of the four baselines and CHIP against ambiguity attack with oracle passport. Acc. (\%) denotes the inference accuracy of the stolen model when the forged passport is present. SDA (\%) and PHA (\%) are measured to verify whether the forged passport passes $\mathcal{V}^D$ and $\mathcal{V}^H$, respectively. ``N/A'' denotes ``not applicable''. Results are measured on BN. Supplementary results for GN are provided in Table S1 of the Supplementary Material.}
    \begin{minipage}{\textwidth}
        \centering
        \renewcommand\arraystretch{1.0}
        \resizebox{0.95\linewidth}{!}{
            \begin{tabular}{l|ccc|ccc|ccc|ccc}
                \Xhline{2\arrayrulewidth}
                \multicolumn{1}{c|}{\multirow{2}{*}{AlexNet}} & \multicolumn{3}{c|}{CIFAR-10} & \multicolumn{3}{c|}{CIFAR-100} & \multicolumn{3}{c|}{Caltech-101} & \multicolumn{3}{c}{Caltech-256} \\ \cline{2-13} 
                \multicolumn{1}{c|}{}                         & Acc.      & SDA      & PHA     & Acc.      & SDA       & PHA     & Acc.       & SDA       & PHA      & Acc.      & SDA       & PHA      \\ \Xhline{2\arrayrulewidth}
                DeepIPR                                       & 89.45    & 100.00   & N/A     & 65.27    & 100.00    & N/A     & 68.51     & 100.00    & N/A      & 41.27    & 100.00    & N/A      \\
                PAN                                           & 89.38    & 100.00   & N/A     & 67.49    & 100.00    & N/A     & 70.87     & 100.00    & N/A      & 41.53    & 100.00    & N/A      \\
                TdN                                           & 89.94    & 100.00   & 50.65   & 67.01    & 100.00    & 46.96   & 68.88     & 100.00    & 50.61    & 42.24    & 100.00    & 52.56    \\
                SteP                                          & 90.91    & 100.00   & 50.56   & 67.52    & 100.00    & 52.43   & 70.54     & 100.00    & 49.22    & 41.78    & 100.00    & 51.30    \\
                CHIP (Ours)                                   & 82.36    & 100.00   & 51.00   & 48.24    & 100.00    & 52.69   & 64.21     & 100.00    & 51.26    & 32.68    & 100.00    & 48.78    \\ \Xhline{2\arrayrulewidth}
            \end{tabular}
        }
    \end{minipage}
    
    \vspace{2mm}
    
    \begin{minipage}{\textwidth}
        \centering
        \renewcommand\arraystretch{1.0}
        \resizebox{0.95\linewidth}{!}{
            \begin{tabular}{l|ccc|ccc|ccc|ccc}
                \Xhline{2\arrayrulewidth}
                \multicolumn{1}{c|}{\multirow{2}{*}{ResNet-18}} & \multicolumn{3}{c|}{CIFAR-10} & \multicolumn{3}{c|}{CIFAR-100} & \multicolumn{3}{c|}{Caltech-101} & \multicolumn{3}{c}{Caltech-256} \\ \cline{2-13} 
                \multicolumn{1}{c|}{}                           & Acc.      & SDA      & PHA     & Acc.      & SDA       & PHA     & Acc.       & SDA       & PHA      & Acc.      & SDA       & PHA      \\ \Xhline{2\arrayrulewidth}
                DeepIPR                                         & 92.78    & 99.99    & N/A     & 70.26    & 100.00    & N/A     & 66.82     & 100.00    & N/A      & 44.54    & 100.00    & N/A      \\
                PAN                                             & 94.42    & 100.00   & N/A     & 75.66    & 100.00    & N/A     & 71.41     & 100.00    & N/A      & 54.09    & 100.00    & N/A      \\
                TdN                                             & 94.50    & 100.00   & 50.90   & 73.75    & 100.00    & 49.34   & 72.97     & 100.00    & 50.55    & 54.16    & 100.00    & 49.38    \\
                SteP                                            & 94.62    & 100.00   & 48.91   & 74.25    & 100.00    & 49.38   & 73.56     & 100.00    & 49.34    & 53.82    & 100.00    & 49.18    \\
                CHIP (Ours)                                     & 94.48    & 100.00   & 50.27   & 65.79    & 100.00    & 50.86   & 68.49     & 100.00    & 48.01    & 45.88    & 100.00    & 51.02    \\ \Xhline{2\arrayrulewidth}
            \end{tabular}
        }
    \end{minipage}
    \vspace{-3mm}
    \label{tab:oracle_attack}
\end{table*}

\textit{Ambiguity attack with oracle passport.} The attacker is assumed to hold the original passport and 30\% of the original training data to create a forged passport. We conducted experiments on the four baselines and our CHIP to evaluate their robustness against this attack. For each forged passport, the inference accuracy (\%), SDA (\%), and PHA (\%) are measured to check whether it passes the performance fidelity test $\mathcal{V}^F$, the signature detection test $\mathcal{V}^D$, and the passport hashing test $\mathcal{V}^H$, respectively.

Experimental results are provided in Table~\ref{tab:oracle_attack}. A 100\% SDA can be achieved across all cases. This finding is not surprising: given the vast parameter space, it is possible to create a forged passport differs significantly from the original one while producing the same signature. Hence, $\mathcal{V}^D$ can be trivially bypassed.

Regarding performance fidelity, forged passports achieve inference accuracy comparable to unprotected models for all four baselines. In contrast, CHIP models exhibit an average accuracy drop of 13.68\% with forged passports, causing false claims to fail $\mathcal{V}^F$ in certain cases. Benefited from the skip connection, significant passport modifications drastically alter affine factors, thus degrading inference performance.

The PHA measured on TdN, SteP, and CHIP remains near 50\%, i.e., as bad as random guessing. For TdN and SteP, it is computationally intractable for the attacker to re-construct the one-way chameleon-hash from the forged passport to the detected signature, thereby making false claims refuted by $\mathcal{V}^H$. Additionally, CHIP provides a stricter test: the attacker must provide both a forged passport passing $\mathcal{V}^H$ and a valid licensor certificate for $\mathcal{V}^L$. Both require finding a collision to the chameleon-hashed signature, which is provably intractable without the secret key $\mathcal{SK}$. Consequently, false claims are rejected for failing $\mathcal{V}^H$ and $\mathcal{V}^L$.

\begin{figure}[t]
    \centering
    \begin{subfigure}{0.45\linewidth}
        \includegraphics[width=\textwidth]{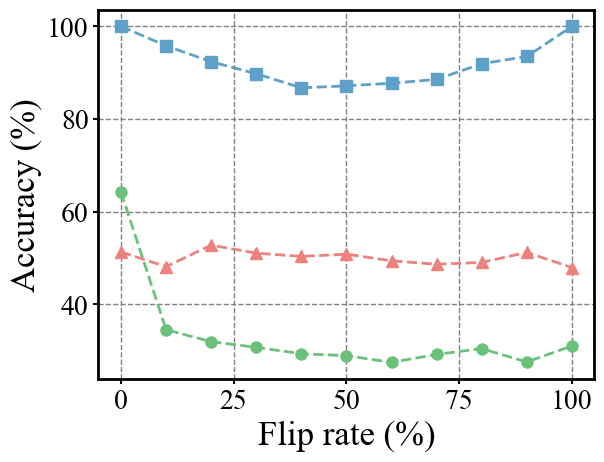}
        \caption{AlexNet}
    \end{subfigure}
    \begin{subfigure}{0.45\linewidth}
        \includegraphics[width=\textwidth]{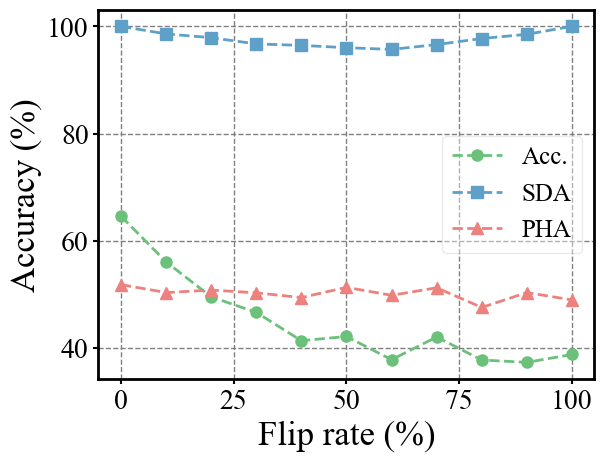}
        \caption{ResNet-18}
    \end{subfigure}

    \caption{Inference accuracy (\%), SDA (\%), and PHA (\%) after ambiguity attacks with oracle passports. The flipping rate ranges from 0\% to 100\%, with a step size of 10\%. The dataset is Caltech-101 and norm type is BN.}
    \vspace{-5mm}
    \label{fig:oracle_flip}
\end{figure}

Beyond forging a passport, the attacker may also attempt to embed a malicious signature that differs from the original. To evaluate this threat, we conduct ambiguity attacks by modifying the original signature with bit-flipping rates ranging from 0\% to 100\%.

Fig.~\ref{fig:oracle_flip} presents the inference accuracy, SDA, and PHA of CHIP models under these attacks. The SDA exhibits a smiling curve across different flip rates. Specifically, as the flip rate increases, the SDA initially decreases gradually until reaching its minimum at 50\% flip rate, then subsequently rises with further increases in flip rate. This behavior is expected, as a flip rate of 50\% represents the most challenging case for malicious signature embedding. On AlexNet, the SDA falls too much to pass $\mathcal{V}^D$ for flip rates between 20\% to 90\%; while on ResNet-18, the SDA maintains above 95\%, indicating successful malicious signature embedding.

However, passing $\mathcal{V}^D$ alone dose not imply a successful removal, replacement or evasion of ownership proof. In fact, such false claims on CHIP models remain detectable. The inference accuracy decreases monotonically with increasing flip rates. Even a modest 10\% flip rate is sufficient to cause a drastic accuracy degradation and fail $\mathcal{V}^F$. Furthermore, the PHA remains near 50\% across different flip rates, demonstrating that these ambiguity attacks consistently fail $\mathcal{V}^H$.

In summary, CHIP is verified to be robust against various ambiguity attacks, regardless of whether the attacker possesses the original passport or additional training data.

\noindent\textbf{Robustness against Removal Attacks}

\begin{figure}[t]
    \centering
    \includegraphics[width=1.0\columnwidth]{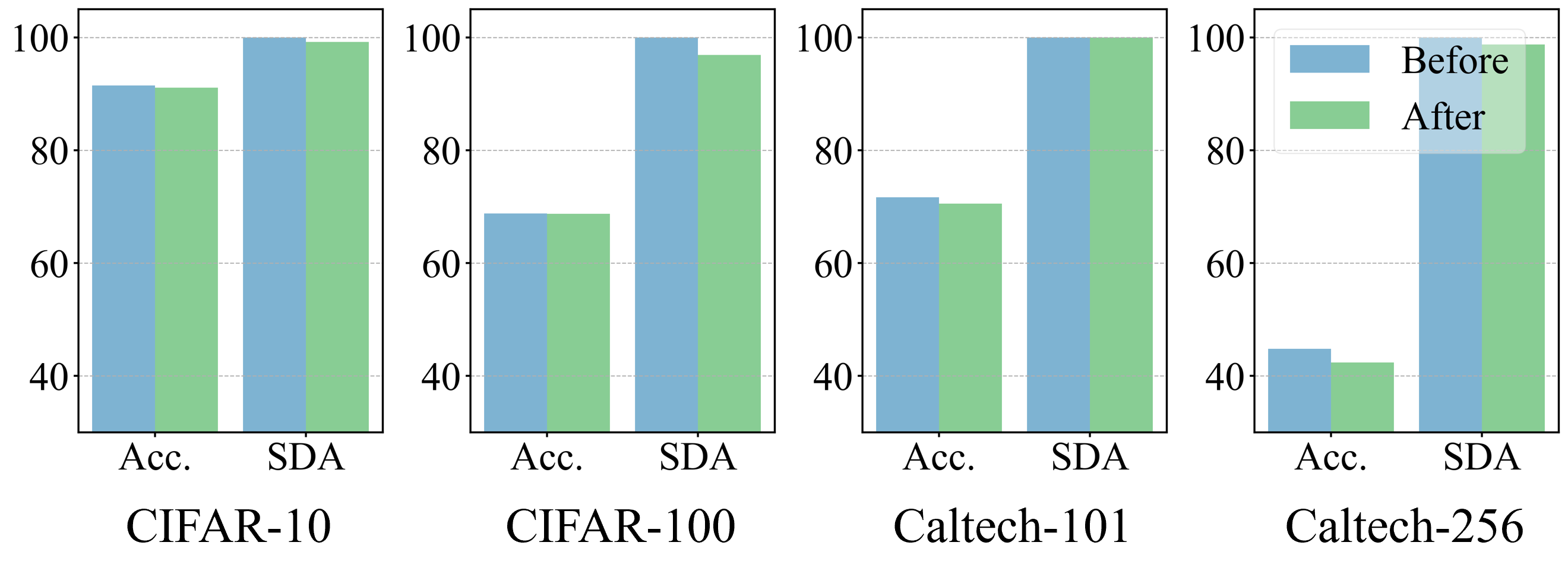}
    \vspace{-2mm}
    \caption{Acc.(\%) and SDA(\%) of CHIP models before and after fine-tuning with 30\% of the original training dataset.}
    \vspace{-5mm}
    \label{fig:fine_tune}
\end{figure}

\textit{Fine-tuning.} To investigate the robustness of CHIP against fine-tuning attack, user models are trained for additional 100 epochs with a small learning rate of 0.001. The attack scenario assumes a realistic setting where the adversary possesses only 30\% of the original training data, otherwise, the attacker can train a new model from scratch rather than fine-tuning a stolen one. Following the re-train all layer (RTAL) configuration~\cite{lukas2022sok}, we first reinitialize the fully-connected layer before optimizing all model parameters.

As shown in Fig.~\ref{fig:fine_tune}, CHIP models display remarkable resilience: even if fine-tuned models can achieve comparable inference accuracy to their original counterparts, the SDA remains consistently above 95\%, underscoring CHIP's effectiveness in maintaining watermark integrity. The embedded watermarks cannot be removed through conventional fine-tuning approaches, even when attackers have partial access to training data and complete model parameter access.

\begin{table*}[t]
    \centering
    \caption{Robustness of the four baseline methods and CHIP against transfer learning attacks. The values outside and inside the brackets represent the optimized model's inference accuracy (\%) and SDA (\%), respectively. The last column reports the mean SDA (\%) of each method.}
    
    \begin{minipage}{\textwidth}
        \centering
        \renewcommand\arraystretch{1.1}
        \resizebox{1.0\linewidth}{!}{
            \begin{tabular}{l|cc|cc|cc|cc|c}
                \Xhline{2\arrayrulewidth}
                \multicolumn{1}{c|}{\multirow{2}{*}{AlexNet}} & \multicolumn{2}{c|}{CIFAR-100 to CIFAR-10} & \multicolumn{2}{c|}{CIFAR-100 to   Caltech-101} & \multicolumn{2}{c|}{Caltech-256 to   CIFAR-10} & \multicolumn{2}{c|}{Caltech-256 to   Caltech-101} & \multirow{2}{*}{\begin{tabular}[c]{@{}c@{}}Mean\\ SDA\end{tabular}} \\ \cline{2-9}
                \multicolumn{1}{c|}{}                         & \multicolumn{1}{c|}{BN}   & GN             & \multicolumn{1}{c|}{BN}     & GN                & \multicolumn{1}{c|}{BN}     & GN               & \multicolumn{1}{c|}{BN}      & GN                 &                                                                     \\ \Xhline{2\arrayrulewidth}
                DeepIPR                                       & 83.96 (100.00)            & 83.54 (98.48)  & 75.93 (99.61)               & 75.93 (96.09)     & 82.89 (100.00)              & 81.13 (99.18)    & 74.75 (100.00)               & 72.43 (98.09)      & 98.93                                                               \\
                PAN                                           & 87.75 (86.33)             & 85.50 (81.12)  & 78.87 (86.68)               & 75.76 (84.59)     & 83.69 (92.19)               & 81.18 (83.64)    & 75.48 (93.71)                & 73.39 (87.20)      & 86.93                                                               \\
                TdN                                           & 87.45 (85.03)             & 85.99 (82.12)  & 78.98 (89.24)               & 76.89 (85.59)     & 83.42 (91.54)               & 81.44 (84.11)    & 74.97 (91.67)                & 72.32 (89.63)      & 87.37                                                               \\
                SteP                                          & 87.57 (81.03)             & 85.70 (84.77)  & 78.76 (79.60)               & 77.10 (87.24)     & 84.10 (91.36)               & 81.33 (87.50)    & 74.69 (93.88)                & 73.22 (93.83)      & 87.40                                                               \\
                CHIP (Ours)                                   & 86.88 (95.88)             & 85.28 (97.92)  & 77.46 (99.48)               & 75.31 (99.87)     & 83.87 (95.10)               & 80.86 (96.53)    & 74.12 (99.70)                & 71.07 (99.61)      & 98.01                                                               \\ \Xhline{2\arrayrulewidth}
            \end{tabular}
        }
    \end{minipage}
    
    \vspace{2mm}
    
    \begin{minipage}{\textwidth}
        \centering
        \renewcommand\arraystretch{1.1}
        \resizebox{1.0\linewidth}{!}{
            \begin{tabular}{l|cc|cc|cc|cc|c}
                \Xhline{2\arrayrulewidth}
                \multicolumn{1}{c|}{\multirow{2}{*}{ResNet-18}} & \multicolumn{2}{c|}{CIFAR-100 to CIFAR-10} & \multicolumn{2}{c|}{CIFAR-100 to   Caltech-101} & \multicolumn{2}{c|}{Caltech-256 to   CIFAR-10} & \multicolumn{2}{c|}{Caltech-256 to   Caltech-101} & \multirow{2}{*}{\begin{tabular}[c]{@{}c@{}}Mean\\ SDA\end{tabular}} \\ \cline{2-9}
                \multicolumn{1}{c|}{}                           & \multicolumn{1}{c|}{BN}   & GN             & \multicolumn{1}{c|}{BN}     & GN                & \multicolumn{1}{c|}{BN}     & GN               & \multicolumn{1}{c|}{BN}      & GN                 &                                                                     \\ \Xhline{2\arrayrulewidth}
                DeepIPR                                         & 88.59 (81.45)             & 84.82 (88.87)  & 78.14 (79.53)               & 71.69 (76.99)     & 84.26 (88.40)               & 79.68 (86.02)    & 74.52 (91.13)                & 69.38 (75.70)      & 83.51                                                               \\
                PAN                                             & 91.19 (83.79)             & 88.99 (91.76)  & 80.79 (82.07)               & 75.82 (88.32)     & 89.06 (90.78)               & 84.54 (86.05)    & 79.89 (88.75)                & 73.56 (82.77)      & 86.79                                                               \\
                TdN                                             & 90.98 (80.78)             & 89.13 (91.56)  & 80.79 (79.53)               & 75.48 (91.31)     & 89.21 (91.68)               & 84.33 (90.78)    & 78.47 (89.26)                & 73.11 (88.87)      & 87.97                                                               \\
                SteP                                            & 90.86 (80.59)             & 88.97 (93.01)  & 80.79 (78.13)               & 75.25 (91.64)     & 89.55 (88.20)               & 83.64 (91.37)    & 79.60 (88.71)                & 71.69 (90.66)      & 87.79                                                               \\
                CHIP (Ours)                                     & 90.68 (99.18)             & 89.11 (99.38)  & 79.44 (99.65)               & 73.90 (99.02)     & 88.51 (98.20)               & 84.17 (96.91)    & 76.27 (99.49)                & 69.83 (99.77)      & 98.95                                                               \\ \Xhline{2\arrayrulewidth}
            \end{tabular}
        }
    \end{minipage}
    \vspace{-3mm}
    \label{tab:transfer_learning}
    
\end{table*}

\begin{figure}[t]
    \centering
    \begin{subfigure}{0.45\linewidth}
        \includegraphics[width=\textwidth]{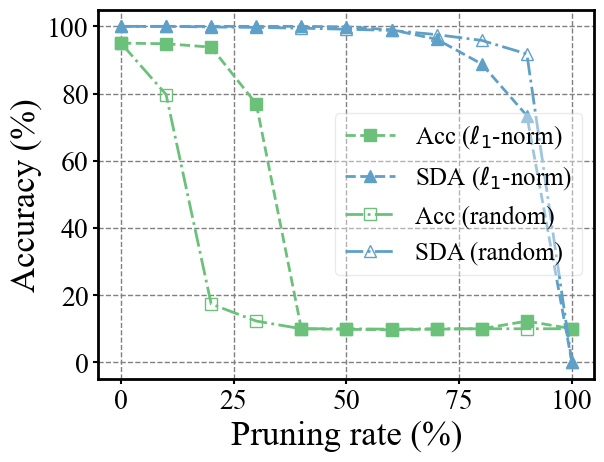}
        \caption{BN}
        \label{fig:pruning_bn}
    \end{subfigure}
    \begin{subfigure}{0.45\linewidth}
        \includegraphics[width=\textwidth]{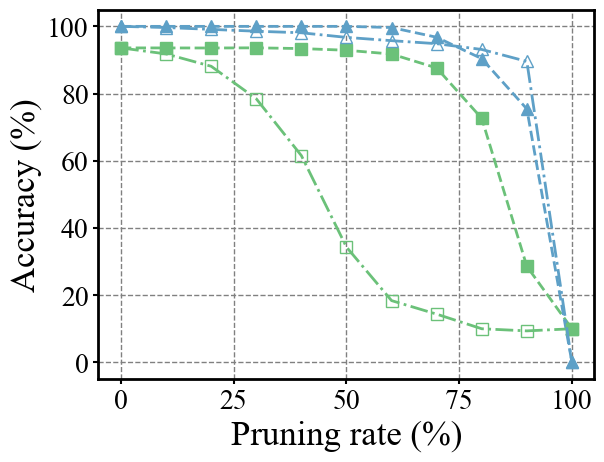}
        \caption{GN}
        \label{fig:pruning_gn}
    \end{subfigure}

    \vspace{-2mm}
    \caption{Inference accuracy (\%) and SDA (\%) after random pruning or $\ell_1$-norm pruning. Evaluations are conducted on ResNet-18 models trained on CIFAR-10.}
    \vspace{-5mm}
    \label{fig:pruning}

\end{figure}

\textit{Transfer learning.} An adversary may attempt to remove an embedded watermark by training a stolen model on a different target dataset. To evaluate the robustness of CHIP against such attacks, we conduct experiments under four transfer learning scenarios: CIFAR-100 to CIFAR-10, CIFAR-100 to Caltech-101, Caltech-256 to CIFAR-10, and Caltech-256 to Caltech-101. In each case, the stolen model is initially trained on the first dataset and then undergoes  transfer learning on the second dataset. The fully-connected layer is first reinitialized to match the class number of the new dataset, and then all weights are trained for 100 epochs with a constant learning rate of 0.001. \textbf{Note}: The attacker can only perform transfer learning on the distributed model that has only one norm branch, as the dual-branch master model is kept private by the model owner. In other words, for the four baselines, the passport-free models are optimized; whereas for CHIP, the passport-aware model are tuned.

Table~\ref{tab:transfer_learning} presents the evaluation results. On AlexNet, DeepIPR and CHIP maintains a high SDA across all cases after transfer learning. Even though all weights are involved in the transfer learning process, the embedded signature can still be successfully extracted from the optimized model with a high SDA of over 98\%. In contrast, the other three baseline methods suffer significant SDA degradation (average 13\% drop), rendering their ownership claims unreliable as the error rates exceed the predefined threshold $\tau_{\text{error}}$. On ResNet-18, DeepIPR attains the lowest averaged SDA of only 83.51\%. The other three baseline methods still maintain an averaged SDA close to 87\%, which is insufficient for reliable ownership verification. CHIP outperforms the four baseline methods with an apparently higher averaged SDA of 98.95\%, successfully passing the signature detection test $\mathcal{V}^D$ and enabling unambiguous ownership attestation.

\textit{Weight pruning.} Following~\cite{cui2024steganographic}, two pruning strategies are considered: random pruning and $\ell_1$-norm pruning. The user model is globally pruned with a pruning rate ranging from 0\% to 100\%, and a step size of 10\%.

Fig.~\ref{fig:pruning} plots the inference accuracy and SDA of CHIP models pruned by the two strategies. It shows that the inference accuracy declines more slowly with $\ell_1$-norm pruning than with random pruning. This is because $\ell_1$-norm pruning tends to eliminate less significant weights with small magnitudes before crucial weights with larger magnitude to avoid performance degradation. The SDA drops much later and more gently than the inference accuracies on pruning, regardless of the pruning strategy. At 80\% pruning rate, the SDA stays above 80\% while the inference accuracies have dropped to below 20\% and 75\% for BN (Fig.~\ref{fig:pruning_bn}) and GN (Fig.~\ref{fig:pruning_gn}), respectively. More supplementary results are presented in Fig. S4 and Fig. S5 of the Supplementary Material. These results corroborate that the attacker cannot remove the embedded signature without significantly degrading the inference performance.

\begin{table*}[t]
    \centering
    \caption{Inference accuracy (\%) of CHIP models trained with different target norm layers.}
    
    \begin{minipage}{\textwidth}
        \centering
        \renewcommand\arraystretch{1.1}
        \resizebox{1.0\linewidth}{!}{
            \begin{tabular}{c|cc|cc|cc|cc|c}
                \Xhline{2\arrayrulewidth}
                \multicolumn{1}{c|}{\multirow{2}{*}{\ AlexNet\ }} & \multicolumn{2}{c|}{CIFAR-10}           & \multicolumn{2}{c|}{CIFAR-100}          & \multicolumn{2}{c|}{Caltech-101}        & \multicolumn{2}{c|}{Caltech-256}        & \multirow{2}{*}{Mean} \\ \cline{2-9}
                \multicolumn{1}{c|}{}                         & \multicolumn{1}{c|}{BN} & GN            & \multicolumn{1}{c|}{BN} & GN            & \multicolumn{1}{c|}{BN} & GN            & \multicolumn{1}{c|}{BN} & GN            &                       \\ \Xhline{2\arrayrulewidth}
                clean                                         & 91.09                   & 89.92         & 68.79                   & 65.05         & 72.20                   & 69.21         & 44.15                   & 41.88         & 67.79                 \\
                \uppercase\expandafter{\romannumeral1}        & 91.23 / 91.23           & 90.00 / 90.01 & 68.21 / 68.20           & 64.17 / 64.18 & 71.81 / 71.81           & 70.11 / 70.11 & 43.61 / 43.64           & 40.51 / 40.51 & 67.46 / 67.46         \\
                \uppercase\expandafter{\romannumeral2}        & 91.45 / 91.48           & 90.07 / 90.05 & 68.77 / 68.78           & 64.37 / 64.38 & 71.69 / 71.69           & 68.93 / 68.93 & 44.80 / 44.82           & 41.00 / 40.98 & 67.64 / 67.64         \\
                \uppercase\expandafter{\romannumeral3}        & 90.91 / 90.93           & 89.71 / 89.69 & 68.37 / 68.39           & 65.45 / 65.49 & 70.23 / 70.23           & 68.02 / 68.02 & 44.06 / 44.02           & 40.60 / 40.62 & 67.17 / 67.17         \\ \Xhline{2\arrayrulewidth}
            \end{tabular}
        }
    \end{minipage}
    
    \vspace{2mm}
    
    \begin{minipage}{\textwidth}
        \centering
        \renewcommand\arraystretch{1.1}
        \resizebox{1.0\linewidth}{!}{
            \begin{tabular}{c|cc|cc|cc|cc|c}
                \Xhline{2\arrayrulewidth}
                \multicolumn{1}{c|}{\multirow{2}{*}{ResNet-18}} & \multicolumn{2}{c|}{CIFAR-10}           & \multicolumn{2}{c|}{CIFAR-100}          & \multicolumn{2}{c|}{Caltech-101}        & \multicolumn{2}{c|}{Caltech-256}        & \multirow{2}{*}{Mean} \\ \cline{2-9}
                \multicolumn{1}{c|}{}                           & \multicolumn{1}{c|}{BN} & GN            & \multicolumn{1}{c|}{BN} & GN            & \multicolumn{1}{c|}{BN} & GN            & \multicolumn{1}{c|}{BN} & GN            &                       \\ \Xhline{2\arrayrulewidth}
                clean                                           & 95.00                   & 93.48         & 76.39                   & 72.16         & 70.68                   & 66.67         & 53.73                   & 45.38         & 71.69                 \\
                \uppercase\expandafter{\romannumeral1}          & 94.80 / 94.79           & 93.51 / 93.51 & 76.64 / 76.64           & 70.91 / 70.91 & 72.54 / 72.60           & 67.74 / 67.68 & 55.04 / 55.07           & 44.90 / 44.93 & 72.01 / 72.02         \\
                \uppercase\expandafter{\romannumeral2}          & 94.67 / 94.68           & 93.52 / 93.52 & 77.23 / 77.22           & 72.54 / 72.87 & 73.28 / 73.33           & 69.89 / 69.89 & 55.53 / 55.63           & 45.41 / 45.46 & 72.76 / 72.83         \\
                \uppercase\expandafter{\romannumeral3}          & 94.83 / 94.86           & 93.93 / 93.93 & 77.23 / 77.17           & 73.50 / 73.54 & 73.62 / 73.67           & 69.21 / 69.27 & 54.51 / 54.46           & 46.78 / 46.75 & 72.95 / 72.96         \\ \Xhline{2\arrayrulewidth}
            \end{tabular}
        }
    \end{minipage}
    \vspace{-3mm}
    \label{tab:target_layers}
    
\end{table*}

\subsection{Cloud Application}

To simulate the deployment of CHIP in the MLaaS scenario, we generate 100 genuine tuples (i.e., $\{p^j_u, r^j_u\}^{100}_{j=1}$), each distributed to a unique end-user as an identity token. Additionally, 300 forged tuples are created to simulate three types of adversarial attacks, including:
\begin{itemize}
    \item Certificate forging attack: 100 tuples with valid user passports but randomized licensee certificates;
    \item Passport forging attack: 100 tuples with valid licensee certificates but randomized passports;
    \item Brute-force attack: 100 tuples with both randomized user passports and licensee certificates.
\end{itemize}

The 400 tuples are used to initiate 400 API calls to a remote ML hosted on a cloud server. For each request, the system computes the chameleon hash value from the submitted passport and licensee certificate, and compares it against the signature to verify authenticity. The evaluation results show 100 true positives and 300 true negatives, with zero false positives or false negatives. Consequently, the system achieves flawless classification performance, attaining 100\% precision, recall, and F1 score. This empirically validates CHIP's provable security against all three attack types in the simulated MLaaS deployment. The collision resistance property of chameleon hash guarantees that attackers cannot forge valid tuples without the secret key. Additionally, CHIP optimizes verification efficiency by requiring only a single chameleon hash computation per API request, ensuring scalability for large-scale API authentication.

\subsection{Further Analysis} \label{sec:further_analysis}

\noindent\textbf{Ablation studies}

\textit{Effect of the target norm layers.} We investigate the impact of target norm layers by training CHIP models with varying passport embedding configurations. For AlexNet, we evaluate three configurations: (\uppercase\expandafter{\romannumeral1}) only the last norm layer, (\uppercase\expandafter{\romannumeral2}) the last three norm layers (default), and (\uppercase\expandafter{\romannumeral3}) all five norm layers. For ResNet-18, we test norm layers in: (\uppercase\expandafter{\romannumeral1}) ``layer4'' (default), (\uppercase\expandafter{\romannumeral2}) ``layer3 + layer4'', and (\uppercase\expandafter{\romannumeral3}) ``layer2 + layer3 + layer4''.

As shown in Table~\ref{tab:target_layers}, CHIP models consistently achieve high inference accuracy comparable to those of the unprotected models, regardless of the configurations of target norm layers. Hence, the owner can freely select any subsets of norm layers as passport layers.

\begin{figure}[t]
    \centering
    \begin{subfigure}{0.45\linewidth}
        \includegraphics[width=\textwidth]{figs/active_control/resnet_cifar10_v4_l4/bn/acc.png}
        \caption{with (w)}
        \label{fig:skip_with}
    \end{subfigure}
    \begin{subfigure}{0.45\linewidth}
        \includegraphics[width=\textwidth]{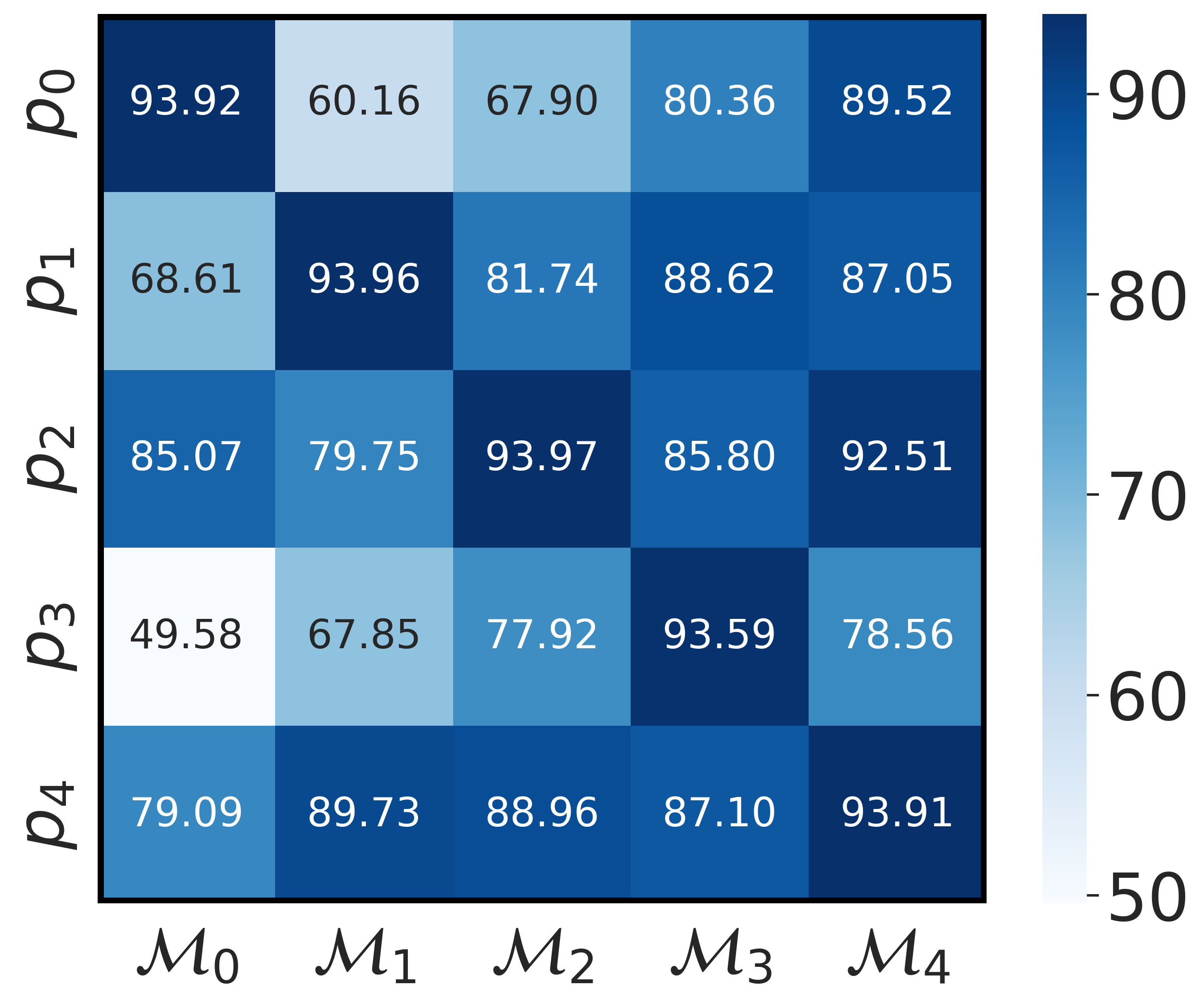}
        \caption{without (w/o)}
        \label{fig:skip_without}
    \end{subfigure}
    \caption{Inference accuracy confusion matrices measured on CHIP models (a) with and (b) without skip connections. Experiments are conducted on ``ResNet-18\_CIFAR-10\_BN''.}
    \vspace{-5mm}
\end{figure}

\textit{Effect of the skip connection.} As discussed in~\ref{sec:stage_1}, the skip connection from the passport to affine factors is critical for effective active control. To evaluate its importance, we train a CHIP model without this skip connection and derive five user models from it. Fig.~\ref{fig:skip_with} demonstrates that models with skip connections perform properly only with their designated paired passports. Without the skip connection, however, non-diagonal elements in the confusion matrix have their values drawn closer to diagonal elements (Fig.~\ref{fig:skip_without}), indicating a failure of active control. For instance, user passport $p_2$ achieves 92.51\% inference accuracy on mismatched user model $\mathcal{M}_4$. This evidence confirms that skip connections make affine factors highly dependent on the passport, ensuring each user model operates exclusively with its designated paired passport.

\begin{table}[t]
    \caption{Inference accuracy (\%) of passport-free/passport-aware models across four graph classification datasets. The first row ``clean'' represents unprotected GIN models without passport layers.}
    \centering
    \renewcommand\arraystretch{1.1}
    \resizebox{0.95\linewidth}{!}{
        \begin{tabular}{l|cccc}
            \Xhline{2\arrayrulewidth}
            \multicolumn{1}{c|}{GIN} & IMDB-B        & COLLAB        & NCI1          & AIDS          \\ \Xhline{2\arrayrulewidth}
            clean                    & 74.00         & 81.00         & 81.27         & 98.83         \\
            CHIP                     & 73.67 / 73.67 & 81.40 / 81.40 & 80.78 / 80.78 & 99.17 / 99.17 \\ \Xhline{2\arrayrulewidth}
        \end{tabular}
    }
    \vspace{-3mm}
    \label{tab:graph}
\end{table}

\noindent\textbf{CHIP for graph classification}

We also evaluated CHIP's performance generalizability beyond image classification by applying it to Graph Isomorphism Networks (GINs)~\cite{xu2019powerful} for graph classification tasks. Protected GINs are trained on four datasets (IMDB-Binary (IMDB-B), COLLAB, NCI1, and AIDS~\cite{rossi2015network}) for 200 epochs, with a 7:3 train-test split ratio. The initial learning rate is set to 0.01, and reduced by half every 50 epochs.

As shown in Table~\ref{tab:graph}, CHIP models achieve comparable inference accuracy to clean models, demonstrating seamless integration into GINs without performance degradation. Owing to the balance loss $\mathcal{L}_\text{bal}$, there is no performance deviation between the two branches across all datasets. Meanwhile, all CHIP models attain a 100\% SDA, confirming successful watermark embedding. We also conduct experiments to assess the active control of CHIP on graph classification tasks, and the results are provided in Fig. S3 of the Supplementary Material. Similar to image classification, each GIN user model operates correctly only with its designated paired passport, and distinct passports exhibit low cosine similarity.

In summary, CHIP's proven effectiveness in watermarking GINs with multi-user access control for graph classification tasks validates its generalizability across diverse ML tasks.

% \begin{figure}[t]
%     \centering
%     \begin{subfigure}{0.45\linewidth}
%         \includegraphics[width=\textwidth]{figs/statistics/resnet_caltech-101_bn.png}
%         \caption{BN}
%         \label{fig:statistics_bn}
%     \end{subfigure}
%     \begin{subfigure}{0.45\linewidth}
%         \includegraphics[width=\textwidth]{figs/statistics/resnet_caltech-101_gn.png}
%         \caption{GN}
%         \label{fig:statistics_gn}
%     \end{subfigure}
%     \caption{Parameter distributions of clean and CHIP ResNet-18 models trained on Caltech-101.}
%     \vspace{-3mm}
%     \label{fig:statistics}
% \end{figure}

% \noindent\textbf{Parameter distribution}

% Statistical distribution differences in model parameters between unprotected and protected models may leak crucial layers or weights contributions to watermarking and active control. Fig.~\ref{fig:statistics} compares the parameter distributions of clean and CHIP models for both BN and GN. Since the two histograms have similar shape and nearly overlap, it can be concluded that statistical analysis cannot be exploited to compromise CHIP.

\noindent\textbf{Complexity}

Table~\ref{tab:time_complexity} compares the training and inference time of clean and CHIP models measured on an NVIDIA A100 GPU. It shows that training a master CHIP model requires approximately 3.71$\times$ (AlexNet) and 3.35$\times$ (ResNet-18) more time than training an unprotected clean model. During inference, the passport-aware branch introduces additional computational overhead for processing passports and generating affine factors, resulting in average time complexities of 1.33$\times$ (AlexNet) and 1.32$\times$ (ResNet-18) compared to clean models.

Security enhancements inevitably involve trade-offs. Notably, the training phase, typically a one-time process managed by the model owner, can accommodate higher computational costs for long-term security benefits. Moreover, the current implementation presents opportunities for further optimization that could reduce training time. The modest increase in inference time remains acceptable even for real-time applications, making CHIP practical for offline deployment scenarios. This reasonable computational overhead represents a worthwhile investment for robust IP protection.
% \footnote{Our project builds upon the official implementation of DeepIPR (https://github.com/kamwoh/DeepIPR), which trains protected models by processing data twice: first for the passport-free branch and then for the passport-aware branch. This process could be optimized by eliminating redundant computation of shared features between branches.}

More importantly, CHIP allows efficient generation of multiple user triplets through trapdoor collision. Table~\ref{tab:time_collision} shows that creating \textbf{five} distinct user triplets only requires several minutes, which is significantly faster than training a single watermarked model from scratch. Moreover, this data-free process exhibits network-dependent but dataset-independent complexity, as user model fidelity is guaranteed by minimizing the balance loss $\mathcal{L}_\text{bal}$ to align affine factors with the master model, rather than through traditional data-driven training. Consequently, CHIP provides excellent flexibility and efficiency to create multiple user triplets, making it particularly suitable for scalable offline distribution scenarios where multiple customized models need to be efficiently generated and distributed for traceability.

\begin{table}[t]
    \caption{Training (T) and inference (I) time of clean and CHIP models. The values are in second/epoch.}
    \centering
    \renewcommand\arraystretch{1.2}
    \resizebox{1.0\linewidth}{!}{
        \begin{tabular}{c|l|cc|cc|cc|cc}
            \Xhline{2\arrayrulewidth}
            \multirow{2}{*}{Model}     & \multicolumn{1}{c|}{\multirow{2}{*}{Type}} & \multicolumn{2}{c|}{CIFAR-10} & \multicolumn{2}{c|}{CIFAR-100} & \multicolumn{2}{c|}{Caltech-101} & \multicolumn{2}{c}{Caltech-256}  \\ \cline{3-10} 
                                       & \multicolumn{1}{c|}{}                      & T              & I            & T              & I             & T               & I              & T               & I              \\ \Xhline{2\arrayrulewidth}
            \multirow{2}{*}{AlexNet}   & clean                                      & 4.02           & 0.42         & 4.02           & 0.41          & 0.74            & 0.26           & 2.03            & 0.35           \\
                                       & CHIP                                       & 15.13          & 0.59         & 15.28          & 0.59          & 2.27            & 0.29           & 7.35            & 0.45           \\ \hline
            \multirow{2}{*}{ResNet-18} & clean                                      & 9.47           & 0.61         & 9.39           & 0.63          & 1.49            & 0.29           & 4.59            & 0.46           \\
                                       & CHIP                                       & 31.81          & 0.93         & 31.93          & 0.67          & 4.56            & 0.37           & 15.24           & 0.67           \\ \Xhline{2\arrayrulewidth}
        \end{tabular}
    }
    \vspace{-1mm}
    \label{tab:time_complexity}
\end{table}

\begin{table}[t]
    \caption{Optimization time (minute) of generating \textbf{five} distinct user triplets based on the master model. The values are measured on an NVIDIA A100 GPU.}
    \centering
    \renewcommand\arraystretch{1.2}
    \resizebox{0.95\linewidth}{!}{
        \begin{tabular}{c|cccc}
            \Xhline{2\arrayrulewidth}
            Model     & CIFAR-10   & CIFAR-100   & Caltech-101   & Caltech-256   \\ \Xhline{2\arrayrulewidth}
            AlexNet   & 6.55       & 6.55        & 6.33          & 6.45          \\
            ResNet-18 & 10.38      & 10.48       & 10.00         & 10.23         \\ \Xhline{2\arrayrulewidth}
        \end{tabular}
    }
    \vspace{-1mm}
    \label{tab:time_collision}
\end{table}

\section{Conclusion}

This paper introduces a novel DNN IP protection method called CHIP. It replaces selected norm layers of the target model with carefully designed passport layers to embed an immutable signature for watermarking and active control simultaneously. The embedded signature is generated by a chameleon hash with the owner passport and a licensor certificate that encode a plaintext copyright text. This design allows the owner to create multiple different user models by efficiently fine tuning the small TLP layers connected locally to each passport-aware branch. Each user model is uniquely bound with a distinct user passport and a licensee certificate generated by the trapdoor collision. A registered user must present the assigned user passport to normally use the distributed model. In the event of IP infringement, the owner can verify ownership of deployed models by presenting the master model's passport-aware branch, along with the owner passport and licensor certificate. The master model can also be deployed for MLaaS for online access by registered users with their assigned user passports and licensee certificates. Comprehensive evaluations across four datasets and two DNN models demonstrate that CHIP can successfully embed a robust signature into the target model without degrading inference performance, and restrict model usage strictly to the specific registered user who holds the correct passport. Our method has also been validated to resist known ambiguity attacks and removal attacks.

\bibliographystyle{IEEEtran}
\bibliography{main}

\end{document}